\documentclass[aps,prl,reprint,twocolumn,superscriptaddress,floatfix,nofootinbib,longbibliography]{revtex4-1}
\usepackage{amsthm}
\usepackage{amsmath,amssymb,color,comment,physics}
\usepackage[makeroom]{cancel}
\usepackage[caption=false]{subfig}
\usepackage{mathrsfs}
\usepackage{graphicx}
\usepackage{subfig}
\usepackage[countmax]{subfloat}
\usepackage[english]{babel}
\usepackage{dsfont}
\usepackage[normalem]{ulem}
\usepackage[bookmarks=true,colorlinks,linkcolor=OrangeRed,urlcolor=NavyBlue,citecolor=RoyalBlue]{hyperref}
\usepackage[dvipsnames]{xcolor}
\usepackage{braket}

\usepackage{color}

\begin{document}

\title{Cavity Induced Topology in Graphene}

\author{Ceren~B.~Dag}
\email{ceren.dag@cfa.harvard.edu}

\affiliation{ITAMP, Harvard-Smithsonian Center for Astrophysics, Cambridge, MA 02138, USA}
\affiliation{ Department of Physics, Harvard University, Cambridge, Massachusetts 02138, USA}

\author{Vasil~Rokaj}
\email{vasil.rokaj@cfa.harvard.edu}

\affiliation{ITAMP, Harvard-Smithsonian Center for Astrophysics, Cambridge, MA 02138, USA}
\affiliation{ Department of Physics, Harvard University, Cambridge, Massachusetts 02138, USA}


\begin{abstract}
Strongly coupling materials to cavity fields can affect their electronic properties altering the phases of matter. We study the monolayer graphene whose electrons are coupled to both left and right circularly polarized photons, and time-reversal symmetry is broken due to a phase shift between the two polarizations. We develop a many-body perturbative theory, and derive cavity mediated electronic interactions. This theory leads to a gap equation which predicts a sizable topological band gap at Dirac nodes in vacuum and when the cavity is prepared in an excited Fock state. Remarkably, band gaps also open in light-matter hybridization points away from the Dirac nodes giving rise to topological photo-electron bands with high Chern numbers. We reveal that the physical mechanism behind this phenomenon lies on the exchange of chiral photons with electronic matter at the hybridization points, and the number and polarization of exchanged photons determine the Chern number. This is a generic microscopic mechanism for the photo-electron band topology. Our theory shows that graphene-based materials, with no need of Floquet engineering and hence protected from the heating effects, host high Chern insulator phases when coupled to chiral cavity fields. 

\end{abstract}

\maketitle

Driving quantum materials by classical light is a mature field of physics \cite{Oka,RevModPhys.93.041002} where one can engineer the band topology of materials \cite{PhysRevB.79.081406,Refael11,Kitagawa11,Gedik13,mciver2020light}. Meanwhile, great progress has been achieved in the manipulation of quantum materials with cavity vacuum fields~\cite{ garcia2021manipulating, schlawin2022cavity, RevModPhys.91.025005,flick2015kohn, hubener2021engineering, ruggenthaler2022understanding, sidler2022perspective, PhysRevB.84.195413,PhysRevB.81.165433,PhysRevB.99.235156,PhysRevB.107.195104,Rokaj2deg,scalari2012ultrastrong, keller2020landau, li2018vacuum, hagenmuller2010ultrastrong, bartolo2018vacuum, rokajtopological2023, bacciconi2023topological}. Notably, modifications in the magneto-transport properties
~\cite{paravicini2019magneto} and the Hall conductivity
~\cite{doi:10.1126/science.abl5818} due to cavity vacuum fluctuations were reported in experiments, as well as a shift in the critical temperature for the metal to insulator transition in 1T-TaS$_2$ \cite{jarc2022cavity}. Recently, Ref.~\cite{hubener2021engineering} discussed an experimentally realizable path to chiral cavities through the Faraday effect \cite{FaradayOnTM,1067270}. Specifically, a magneto-optical material coated mirror would induce a phase shift between the two polarizations of the electromagnetic cavity field \cite{Arikawa:12,chin2013nonreciprocal} where the phase shift is proportional to the applied magnetic field, thickness of the coating and the Verdet constant \cite{carothers2022high}. Such Faraday rotators  \cite{Arikawa:12} and metamaterial coated mirrors \cite{plum2015chiral} were also experimentally demonstrated to selectively absorb one polarization or the other, potentially leading to single-polarization chiral cavities. Alternative to an external magnetic field, spontaneous material magnetism can also be utilized for Faraday effect \cite{Rudner_2019}.

Here we theoretically study a model where a graphene monolayer is coupled to a chiral cavity field with single or two circular polarizations. For the latter, the time-reversal symmetry (TRS) can be broken as a result of an imperfect phase shift with a Faraday mirror, so that one of the polarizations is not eliminated, but only suppressed. In such a setup, what breaks the TRS is the unequal light-matter couplings induced by the two polarizations of the same cavity mode. We formulate a many-body perturbative theory for the continuum Dirac Hamiltonian coupled to light, based on the Schrieffer–Wolff (SW) transformation \cite{PhysRev.97.869, SWolf}, and obtain the cavity mediated electronic interactions. Then, we apply Hartree-Fock mean-field theory (MFT) and show that the cavity mediated interactions break TRS, and hence open a topological gap. Further, we derive the gap equations at finite temperature for a cavity either in vacuum or in a Fock state with low photon number. The perturbative treatment captures the numerically predicted enhancement of the gap with the number of chiral photons when the cavity is prepared in a Fock state \cite{rivera2023creating}. By also deriving a minimally coupled tight-binding (TB) Hamiltonian for this setup and examining the band structure, we show that our results remain valid within the microscopic theory. Hence we find that the single-polarization model \cite{PhysRevB.81.165433,PhysRevB.99.235156,PhysRevB.107.195104} overestimates the Dirac gap in vacuum when Faraday rotation cannot eliminate one of the polarizations. 

A central finding of our work is that TRS breaking also opens topological gaps between the higher energy bands in photo-electron band structure, and away from the Dirac nodes. These topological gaps, being a signature of avoided crossings between strongly coupled electron and photons, contribute nonzero Berry phase to the band wave functions giving rise to higher Chern bands. We unveil the mechanism behind this phenomenon based on the chiral photon exchange processes with matter, and find that the number and polarization of the exchanged photons determine the topology of the photo-electron bands. Our work provides a physically intuitive and generic framework to engineer photo-electron bands with arbitrary Chern numbers, as well as a possible microscopic origin of Floquet topological insulators \cite{PhysRevB.79.081406} with high Chern numbers \cite{PhysRevLett.113.236803}. 

\begin{figure*}
\centering
\includegraphics[width=1.95\columnwidth]{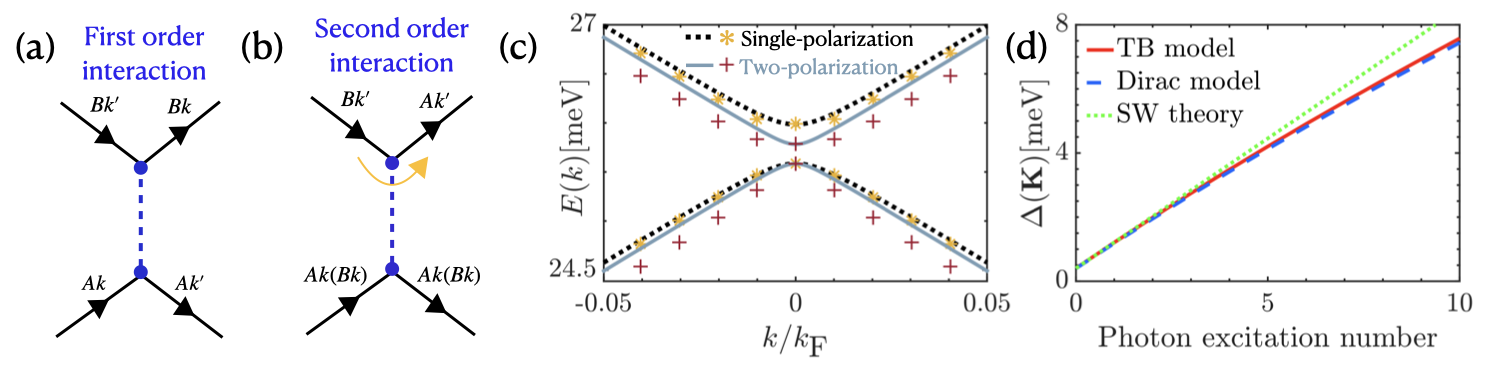}
\caption{The cavity mediated electronic interactions in graphene with strengths proportional to (a) $1/\omega_j$ and (b) $1/\omega_j^2$. (c) The focus on the lowest two bands of two- and single-polarization Dirac models showing the match between analytical SW theory and exact diagonalization (ED) with a truncated photonic Hilbert space of maximum photon number $\langle a_j^{\dagger}a_j\rangle_{\textrm{max}}=4$. Single-polarization bands (dotted-black) are shifted upwards for comparison by $\omega_L/2$. The x-axis is defined in terms of radial distance to the Dirac nodes $k=\sqrt{k_x^2+k_y^2}$ normalized by Fermi momentum $k_{\textrm{F}}=m v_{\textrm{F}}$.
The parameters are $\omega_c=6.28$ THz, $\chi=5\times 10^{-4}$ \cite{paravicini2019magneto} and $m=0.02m_e$. Fermi velocity is found to be $v_{\textrm{F}}=0.21$ a.u. (atomic units) by fitting the band structure of the Dirac model to the TB model. The red-pluses and yellow-stars are the prediction of the analytical SW theory in vacuum. (d) The gap at $\mathbf{K}$ point when the cavity is prepared in a Fock state, increases with the photon number populated in the cavity. SW theory can predict the gap until $\langle  a_R^{\dagger}  a_R\rangle\sim 5$. The ED results with $\langle a_j^{\dagger}a_j\rangle_{\textrm{max}}=4$ on Dirac and TB models match. }
\label{Fig1}
\end{figure*}

\textit{Cavity Mediated Interactions.}~We consider a graphene monolayer described by the continuum Dirac model \cite{bernevig2013topological} placed in a single-mode cavity with frequency $\omega_c$ whose polarizations are in-plane such that they couple to electrons. The effective Hamiltonian around the $\mathbf{K}$ valley reads ($\hbar=1$)~\cite{PhysRevB.99.235156}
\begin{eqnarray}
     \mathcal{H}_{\mathbf{K}}&=& v_F \sum_{\mathbf{k}} \bigg(k_x-A_x+i[k_y-A_y]\bigg)  c_{A\mathbf{k}}^{\dagger} c_{B\mathbf{k}} + \text{h.c.}\notag \\
&+&\sum_{\lambda=R,L}\omega_{\lambda}\left( {a}^{\dagger}_{\lambda} {a}_{\lambda}+\frac{1}{2}\right). \label{fullHamiltonian}
\end{eqnarray}
where the Fermi velocity $v_F=0.21$~a.u.~is found by comparing the band structure of $\mathcal{H}_{\mathbf{K}}$ to that of TB model around the Dirac nodes \cite{supp}. The operators $c_{r\mathbf{k}}$ are fermionic annihilation operators  at $r=A,B$ sublattices with momentum $\mathbf{k}$ obeying $\lbrace c^{\dagger}_{r\mathbf{k}},c_{s\mathbf{k'}} \rbrace = \delta_{\mathbf{k}\mathbf{k}'}\delta_{rs}$. The frequencies of the right- and left-circular polarizations are $\omega_{\lambda}=\sqrt{\omega_c^2+\omega_D^2}$ where $\lambda=R,L$ and the diamagnetic frequency $\omega_D$ stemming from $\mathbf{A}^2$ term, shifts the cavity frequency $\omega_c$ \cite{supp}. The quantized vector potential written in terms of the circular polarizations $\mathbf{e}_{R,L}=(1,\pm \textrm{i})/\sqrt{2}$ is,
\begin{equation}
    {\mathbf{A}}=\sqrt{\frac{1}{\epsilon_0\mathcal{V}2\omega_{R(L)}}}\left[\mathbf{e}_R {a}_L+\mathbf{e}_R {a}^{\dagger}_R+\mathbf{e}_L {a}_R+\mathbf{e}_L {a}^{\dagger}_L\right].\notag
\end{equation}
$\mathcal{V}=\chi \left(2\pi c/ \omega_{c}\right)^3 $ \cite{paravicini2019magneto} is the effective cavity volume with a light concentration parameter $\chi$. Here the operators $[ a_{\lambda},a^{\dagger}_{\lambda'} ] = \delta_{\lambda\lambda'}$ are the circularly polarized photon operators renormalized by the diamagnetic $\mathbf{A}^2$ term originating from the minimally coupled TB Hamiltonian \cite{supp}. Thus the light-matter interaction  Hamiltonian follows as $ \mathcal{H}_{\text{int}}=-v_F \sum_{\mathbf{k}} 
(g_R a_R^{\dagger}+g_L a_L) c_{A\mathbf{k}}^{\dagger} c_{B\mathbf{k}} + \text{h.c.}$ The light-matter coupling amplitudes in terms of the microscopic parameters are obtained to be $g_{\lambda} = \frac{\alpha}{m}\sqrt{2\pi/(\mathcal{V} \hspace{.5mm}\omega_{\lambda}})$ in the TB model derivation \cite{supp}, where $\alpha=2.68$ a.u.~is the lattice distance and $m$ is the effective mass of the electrons subject to crystal potential which should be fixed by the experiment \cite{Zhang_2005}. We set a modest difference between the couplings of the two polarizations, $g_{R} = \sqrt{2}g_L$ originating from the Faraday rotation.

To derive the cavity-mediated interactions we perform SW transformation $H_{\mathbf{K}} = e^{S} \mathcal{H}_{\mathbf{K}} e^{-S}$~\cite{PhysRev.97.869,SWolf}. The light-matter Hamiltonian $\mathcal{H}_{\mathbf{K}}$ is splitted into non-interacting $\mathcal{H}_0$ and interaction part $\mathcal{H}_{\textrm{int}}$, $\mathcal{H}_{\mathbf{K}}=\mathcal{H}_0+\mathcal{H}_{\textrm{int}}$. Then, the operator $S=S_1+S_2+\cdot\cdot$ is constructed perturbatively as an expansion in orders of $1/\omega_{\lambda}$, such that the light-matter interaction is eliminated $[S,\mathcal{H}_0]=-\mathcal{H}_{\textrm{int}}$~\cite{SWolf}. To first order in this expansion we find \cite{supp}
\begin{eqnarray}
    S_1 &=& v_F\sum_{\mathbf{k}} \left(\frac{g_L }{\omega_L} a_L - \frac{g_R }{\omega_R} a_R^{\dagger}  \right) c_{A\mathbf{k}}^{\dagger} c_{B\mathbf{k}}  - \textrm{h.c.}\label{S1op}
\end{eqnarray}
Given the operator $S$, we derive the effective SW Hamiltonian $H_{\mathbf{K}}= \mathcal{H}_0+\frac{1}{2}[S,\mathcal{H}_{\textrm{int}}]$. For a cavity in vacuum this takes the form of
\begin{eqnarray}  
H_{\mathbf{K}} &=& \frac{\omega_R+\omega_L}{2} + v_F \sum_{\mathbf{k}} \bigg[ (k_x+ik_y)  c_{A  \mathbf{k}}^{\dagger} c_{B  \mathbf{k}} \notag\\
&-& \sum_{\mathbf{k}'} \bigg ( \frac{g_R^2}{2\omega_R} c_{B  \mathbf{k}}^{\dagger}c_{A  \mathbf{k}}  c_{A  \mathbf{k}'}^{\dagger}c_{B  \mathbf{k}'} \notag\\
&+& \frac{g_L^2}{2\omega_L} c_{A  \mathbf{k}}^{\dagger}c_{B  \mathbf{k}}  c_{B  \mathbf{k}'}^{\dagger}c_{A  \mathbf{k}'} \bigg)  + \text{h.c.}\bigg] \label{eq:fullEffTwoPolHatKmain}
\end{eqnarray}
The diagrammatic representation of the interactions is given in Fig.~\ref{Fig1}(a). One can obtain the effective Hamiltonian at $\mathbf{K}'$ valley $H_{\mathbf{K'}}$ by exchanging the sublattice indices $A \leftrightarrow B$ and momentum $\mathbf{k} \rightarrow -\mathbf{k}$ in $H_{\mathbf{K}}$. The cavity-mediated interactions break TRS for $g_R \neq g_L$ which we prove below, and estimate the induced gap by MFT whose details are in the SM \cite{supp}. The MFT Hamiltonians read $H^{\textrm{mft}}_{\mathbf{K}} = \sum_{\mathbf{k}}[ v_F'( k_x \sigma_1 + k_y \sigma_2 ) - d_3(\mathbf{k}) \sigma_3 ] + E_0$ and $H^{\textrm{mft}}_{\mathbf{K'}} =   \sum_{\mathbf{k}}[ v_F'(- k_x \sigma_1 + k_y \sigma_2 ) + d_3(\mathbf{k}) \sigma_3 ] + E_0$ at $\mathbf{K}$ and $\mathbf{K'}$ points, respectively, where $\sigma_{1,2,3}$ are the Pauli matrices. Here $v_F'$ is the renormalized Fermi velocity, and $E_0$ is the many-body ground state energy predicted by the MFT, which matches with the band structure results \cite{supp}. Presence of a nonzero $d_3(\mathbf{k})$ in these MFT equations with a different sign means that the TRS is broken. This cavity induced gap shows that both Dirac nodes contribute $\pi$ Berry phase to the wave function, and hence the band gap is topological. We obtain the gap equations for both polarizations to be
\begin{eqnarray}
\Delta_{\lambda}(\mathbf{k}) &=& \frac{g_{\lambda}^2 v_F^2}{2\omega_{\lambda}} \bigg( 1+\Delta_{\lambda}(\mathbf{k})\frac{\tanh (\beta E_{\mathbf{k}}^{\lambda}/2)}{2E_{\mathbf{k}}^{\lambda}} \bigg),\label{gapEqR}
\end{eqnarray}
where $\beta=1/k_BT$ is the inverse temperature, $E_{\mathbf{k}}^{\lambda} =\sqrt{ v_F^{'2}(k_x^2+k_y^2)+\left(\Delta_{\lambda}(\mathbf{k})\right)^2/4}$, and total band gap opening due to interactions is $\Delta(\mathbf{k})\equiv\Delta_R(\mathbf{k})-\Delta_L(\mathbf{k})=2d_3(\mathbf{k})$ . Right at the Dirac nodes and zero temperature, the gap reads $\Delta(\mathbf{0}) = g_R^2v_F^2/\omega_R-g_L^2v_F^2/\omega_L$. Hence in fact, the condition $g_R\neq g_L$ opens a gap. In the limit $T \rightarrow \infty$, the gap reduces to $\Delta(\mathbf{0}) = g_R^2v_F^2/2\omega_R-g_L^2v_F^2/2\omega_L$. The general solution at $T =0$ that is plotted in Fig.~\ref{Fig1}(c) with red pluses, matches with the band structure of the Dirac model in vacuum. The finite-temperature gap is numerically solved in the SM \cite{supp}. 

\begin{figure*}
\centering
\includegraphics[width=1.95\columnwidth]{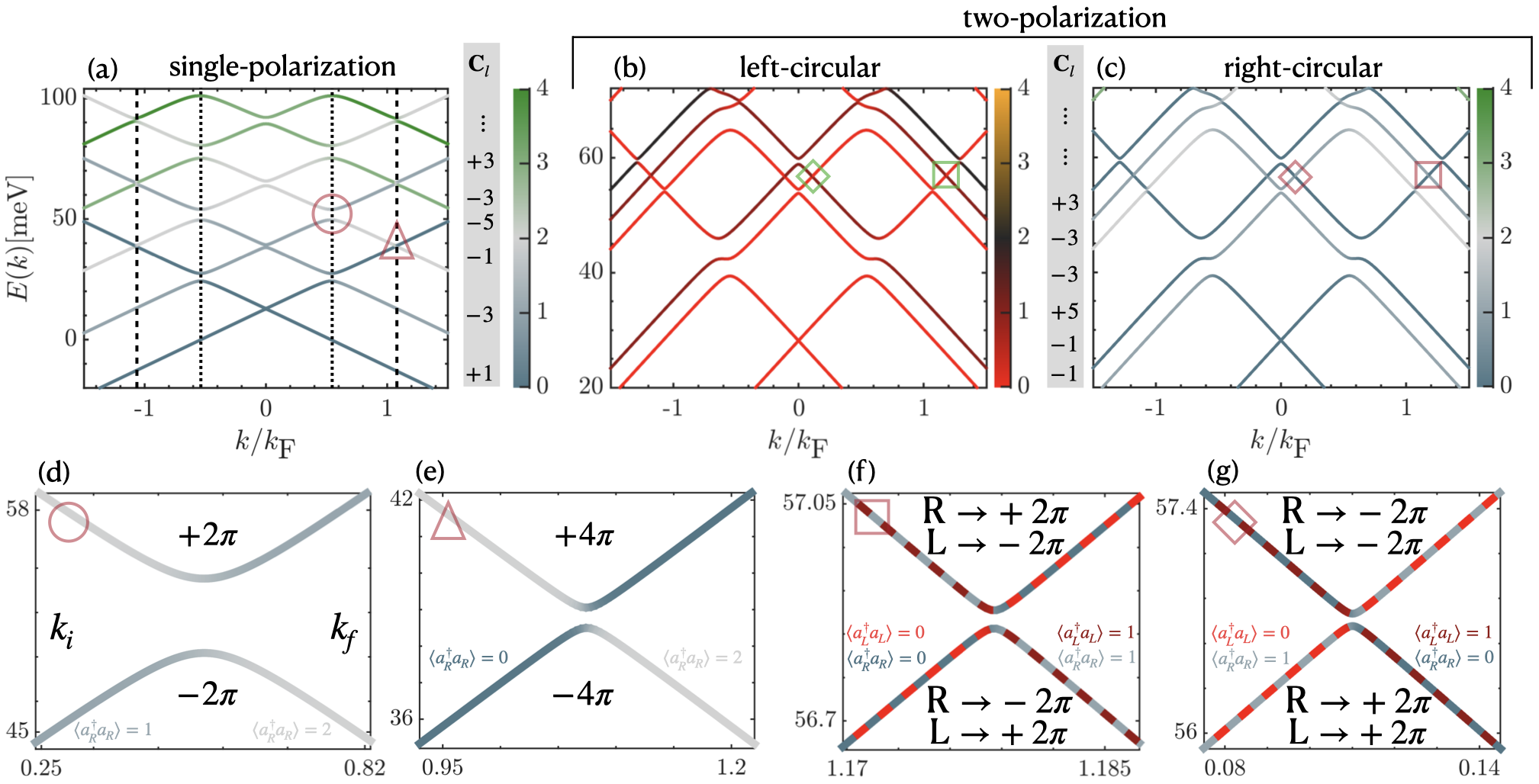}
\caption{The exact diagonalization Dirac bands of graphene around the $\mathbf{K}$ point coupled to a cavity with (a) single polarization and (b-c) two-polarization of cavity frequency $\omega_c=6.28$ THz with color coding denoting the photon populations and $\langle a_j^{\dagger}a_j\rangle_{\textrm{max}}=4$. We use $m=0.02m_e$, $\chi=5\times 10^{-4}$ \cite{paravicini2019magneto} and $\gamma=1.2$ as the free parameters of the theory that have to be fixed by the experiment. The light-red shapes highlight the light-matter avoided crossings with chiral photon exchanges. The Chern numbers for both models are given in boxes under $\mathbf{C}_l$ for band $l$ and calculated in the full BZ with the TB model. (d-g) Focus on chiral photon exchange processes with Berry phases for each band denoted. The photon numbers are written only for the lower bands. Subfigures (d-e) show the role of photon number, whereas (f-g) show also the role of polarization in determining the Berry phase of the photo-electron wave function at a light-matter avoided crossing.  }\label{Fig2}
\end{figure*}

The single-polarization limit can be obtained by taking $g_L=0$ in Eqs.~\eqref{S1op} and \eqref{eq:fullEffTwoPolHatKmain}. Due to the relative simplicity of this limit, we derive the SW Hamiltonian up to the second order in the perturbation theory with the additional transformation term $S_2 = g_R v_F^2/\omega_R^2 \sum_{\mathbf{k}} \left[ a_R(k_x+ik_y)-\textrm{h.c.}  \right] (n_{A  \mathbf{k}}-n_{B  \mathbf{k}})$, and we include higher photon excitations with a cavity prepared in a Fock state, such that $\langle a^{\dagger}_R a_R\rangle \in \mathbb{N}$, and we find
\begin{eqnarray} 
H^{\textrm{sp}}_{\mathbf{K}} &=& v_F \sum_{\mathbf{k}} \bigg(1 -  \frac{v_F^2 g_R^2}{\omega_R^2}   \langle a_R^{\dagger}   a_R \rangle \bigg)  (k_x+ik_y) c_{A  \mathbf{k}}^{\dagger} c_{B  \mathbf{k}} \notag \\
&+&\text{h.c.} +  \omega_R \left(  \langle a_R^{\dagger}  a_R\rangle + \frac{1}{2}\right) -  \frac{g_R^2 v_F }{\omega_R}   \langle a_R^{\dagger}   a_R \rangle \sum_{\mathbf{k}} \notag \\
& &\left(n_{B  \mathbf{k}}-n_{A  \mathbf{k}}\right) -  \frac{g_R^2v_F}{2\omega_R}  \sum_{\mathbf{k}  \mathbf{k}'} \bigg (c_{B  \mathbf{k}}^{\dagger}c_{A  \mathbf{k}}  c_{A  \mathbf{k}'}^{\dagger}c_{B  \mathbf{k}'} \notag \\
&+&  v_F\frac{k_x+ik_y}{\omega_R} (n_{A  \mathbf{k}}-n_{B  \mathbf{k}})c_{A  \mathbf{k}'}^{\dagger}c_{B  \mathbf{k}'}  + \text{h.c.} \bigg). \label{eq:fullEffHatKmain} 
\end{eqnarray}
The interaction induced in the second order with $\propto 1/\omega_R^2$ in Eq.~\eqref{eq:fullEffHatKmain} has a complex amplitude and does not preserve sublattice flavor, as depicted in Fig.~\ref{Fig1}(b). The gap opening introduced in the first order with Eq.~\eqref{gapEqR} is modified by the photon number
\begin{eqnarray}
\Delta_R(\mathbf{k}) &=&\frac{g_R^2 v_F^2}{2\omega_R} \bigg( 1+\Delta_R(\mathbf{k})\frac{\tanh (\beta E_{\mathbf{k}}^R/2)}{2E_{\mathbf{k}}^R} \bigg) \notag\\
&+& 2\frac{g_R^2v_F }{\omega_R}\langle a_R^{\dagger}  a_R\rangle,\notag
\end{eqnarray}
and accompanied with the renormalization of the Fermi velocity in the second order, 
\begin{eqnarray}
v_F' &=& v_F \bigg(1 - \frac{v_F^2 g_R^2}{\omega_R^2} \langle a_R^{\dagger}  a_R \rangle \bigg). \label{gapEqFV}
\end{eqnarray}
In the single-polarization model, $\Delta(\mathbf{k})=\Delta_R(\mathbf{k})$ by definition. At zero temperature, the gap at the Dirac nodes scales as $\Delta(\mathbf{0}) = (2 \langle a^{\dagger}_R a_R \rangle +1) g_R^2v_F^2/\omega_R$ which is compatible with Refs.~\cite{PhysRevB.99.235156,PhysRevB.107.195104} in vacuum. Therefore, populating the cavity does not only increase the topological band gap (Fig.~\ref{Fig1}(d)), it also flattens the bands around the Dirac nodes as is visible in Fig.~\ref{Fig2}(a). The SW theory predicts the gap until the photon excitation number is $\langle  a^{\dagger}_R   a_R\rangle \sim 5$ (Fig.~\ref{Fig1}(d)). Let us note that applying MFT to the second order interaction gives rise to coupled gap equations for $v_F'$ and $\Delta(\mathbf{k})$ whose numerical solutions in generic conditions can be found in the SM \cite{supp}. We plot these solutions in Fig.~\ref{Fig1}(c) in vacuum with yellow stars on the single-polarization Dirac model bands, and see perfect match. Overall, for a split-ring resonator \cite{PhysRevB.90.205309} with $\omega_c = 6.28$ THz cavity frequency ---corresponding to a Hartree energy of $\sim 9.5\times 10^{-4}$ a.u.---, $\chi\sim10^{-4}$ \cite{paravicini2019magneto} and $m=0.007m_e$ \cite{Zhang_2005} where $m_e$ is the bare electron mass, two-polarization model leads to $4.3$meV gap in vacuum, which is overestimated by the single-polarization model, $11.5$meV \cite{supp}. This overestimation is visualized in Fig.~\ref{Fig1}(b) for a set of different parameter values, and what vacuum gap depends on is given in the SM \cite{supp}. These gaps can be measured via transport \cite{Cao_2018}, or angle-resolved photo-emission spectroscopy \cite{RevModPhys.93.025006}. 

\textit{Topological photo-electron bands in graphene.} For the following discussion, we numerically calculate the Berry curvature $F_{l,xy}(k_x,k_y)$ over the full Brillouin zone of the TB models and the Chern number of a band $l$ \cite{Fukui_2005}
\begin{eqnarray}
    C_l=\frac{1}{2\pi} \int_{\textrm{B.Z.}} F_{l,xy}(k_x,k_y)dk_x dk_y.
\end{eqnarray}
The Berry phases at a Dirac node and light-matter avoided crossing are denoted by $\phi_{\textrm{m},l}$ and $\phi_{\textrm{p},l}$, respectively for band $l$. Let us note that all photo-electron Dirac bands plotted in Fig.~\ref{Fig2} are cross-sections cutting through a Dirac node. Hence, the avoided crossings seen symmetrically placed around $\mathbf{K}$ point are two points residing on a continuous loop of hybridizations around $\mathbf{K}$ point \cite{supp}. Therefore, $\phi_{\textrm{p},l}$ counts the Berry phase contribution of all avoided crossings at the same radial distance $k/k_{\textrm{F}}$ to the $\mathbf{K}$ point. The Chern number of the band $l$ is $C_l=\Phi_{l}/\pi$ where $\Phi_{l}$ is the total Berry phase.

The lowest band, $l=1$, has two Dirac nodes each contributing to the winding phase of the wave function $\phi_{\textrm{m},1}=\Phi_{1}=\pi$ leading to a Chern band of $C_1=1$ as numerically confirmed. However, a more significant characteristic of graphene coupled to a chiral cavity is the emergence of the topological light-matter hybrizations reminiscent of topological polaritons \cite{PhysRevX.5.031001}. All higher energy bands enjoy additional Berry phases proportional to the exchanged photon number: the gaps closest to the valleys, dotted-black in Fig.~\ref{Fig2}(a), are 1-photon avoided crossings with 1 chiral photon exchange. This exchange process is enlarged in Fig.~\ref{Fig2}(d). As a result, the second band gains $\phi_{\textrm{p},2}=-2\pi$ phase at these 1-photon avoided-crossings, leading to $\Phi_{2}=-3\pi$ total phase together with the $\phi_{\textrm{m},2}=-\pi$ at $\mathbf{K}$ valley giving rise to $C_2=-3$ as numerically confirmed. The 2-photon avoided crossings depicted with dashed-black in Fig.~\ref{Fig2}(a)-(e), carry $-4\pi$ phase for $l=3$. Therefore, each higher energy band has an additional loop of light-matter avoided crossings with a phase proportional to $\phi_{\textrm{p},l} = -2\pi (\langle a_R^{\dagger} a_R \rangle^{k_f}_{l} - \langle a_R^{\dagger} a_R \rangle^{k_i}_{l})$ contributing to $\Phi_{l}$, and hence to the Chern number of the band $l$, where $|k_f| > |k_i|$ is set as the convention. Berry curvature supports this mechanism, see SM \cite{supp}. 

Polarization of the exchanged photons also affects the Berry curvature and the Chern number of the photo-electron band. Here we consider a two-polarization model and adopt an alternative mechanism to break TRS through a frequency splitting between two polarizations $\omega_R \neq \omega_L$. This model might be realized either via Zeeman splitting \cite{PhysRevB.81.165433,suárezforero2023chiral} or with two Faraday mirrors which selectively absorb one of the polarizations of two cavity modes $\omega_{1}$ and $\omega_{2}$. We parametrize the frequency difference in terms of $\omega_2 = \gamma \omega_1$ where $\gamma \in \mathbb{R}^+$ resulting in  $\omega_{R(L)}=\sqrt{\omega_{1(2)}^2+\omega_D^2}$.  One of our central results it that the Berry phase at a light-matter hybridization can be predicted by 
\begin{eqnarray}
\frac{\phi_{\textrm{p},l}}{2\pi}= \langle a^{\dagger}_L a_L\rangle^{k_f}_{l}-\langle a^{\dagger}_La_L \rangle^{k_i}_{l} - \big(\langle a^{\dagger}_Ra_R \rangle^{k_f}_{l}-\langle a^{\dagger}_Ra_R \rangle^{k_i}_{l}\big). 
\end{eqnarray}
This gives rise to four different cases in the prediction of the Berry phases at the avoided crossings, two of which are enlarged in Fig.~\ref{Fig2}(f)-(g). As depicted with a square in Figs.~\ref{Fig2}(b)-(c), at an avoided crossing between $l=5$ and $l=6$ two photons with opposite chiralities are exchanged with matter leading to a zero Berry phase $\phi_{\textrm{p},5}=\phi_{\textrm{p},6}=0$, and hence a trivial gap. Depicted with a rhombus in Figs.~\ref{Fig2}(b)-(c), at an avoided crossing between $l=4$ and $l=5$ a photon changes chirality through the interactions with matter leading to $\phi_{\textrm{p},4}=2\pi[(1-0)-(0-1)]=4\pi$ and $\phi_{\textrm{p},5}=-4\pi$ Berry phase. In a simpler avoided crossing where a photon of fixed polarization is not exchanged at all, e.g.,~a left-circularly polarized photon for band $l=2$ in Figs.~\ref{Fig2}(b)-(c), Berry phase is contributed only by an exchange between a right-circularly polarized photon and matter, thus reproducing the single-polarization limit.

Therefore, Chern insulator phases with higher Chern numbers can be engineered by utilizing chiral photonic fields. This mechanism seems very general, and not restricted to graphene. For instance, high Chern numbers were reported in transition metal dichalcogenides coupled to single-polarization cavity field \cite{nguyen2023electronphoton}. Furthermore, the topological bands of the bulk suggests chiral edge modes with electron-photon localized states \cite{PhysRevX.5.031001}. Our observation of high Chern numbers might also suggest larger photo-electron currents at the edges, or the domain walls, of the sample which could lend itself to device applications.

\textit{Discussion and Outlook.}---We studied graphene subject to a chiral cavity field where TRS is broken through unequal coupling of left- and right-circularly polarized photons to the electrons.  Hence, we find a sizable Dirac node splitting even in vacuum. The band gap increases when the cavity is populated with photons, facilitating its experimental measurement. Our analytical theory reveals chiral cavity-mediated electronic interactions in graphene. Understanding the competition of the cavity-mediated interactions with Coulomb interactions is an exciting future direction. This theory can also be applied to moir\'{e} materials~\cite{Moiremarvels} coupled to cavities which can guide the exploration on how enhanced vacuum fluctuations affect strongly correlated electron systems. 
Most importantly, the light-matter entanglement in the vicinity of the avoided crossings induces a nonzero Berry phase to the photo-electron wave function, leading to a rich topology based on the exchange processes of chiral photons with electronic matter. Our theory provides insights and intuition on the nature of topological photo-electron bands suggesting a microscopic mechanism underlying high Chern numbers in periodically-driven systems, and establishes a connection between Floquet and cavity engineering of materials.

\textit{Acknowledgments.}---We are grateful to Ashvin Vishwanath for many fruitful and guiding discussions on this work. Authors additionally thank Tilman Esslinger, Mohammad Hafezi, P. Myles Eugenio, Dan Parker, Pavel Volkov and Jie Wang for stimulating discussions, Oriana Diessel and Volker Karle for helpful comments on the paper. The authors acknowledge support from the NSF through a grant for ITAMP at Harvard University.

\bibliographystyle{apsrev4-1}

%

\onecolumngrid
\newpage

\setcounter{equation}{0}
\setcounter{figure}{0}
\setcounter{table}{0}
\setcounter{page}{1}
\makeatletter
\renewcommand{\theequation}{S\arabic{equation}}
\renewcommand{\thefigure}{S\arabic{figure}}
\renewcommand{\bibnumfmt}[1]{[S#1]}
\setcounter{secnumdepth}{1}
\setcounter{secnumdepth}{2}

\begin{center}
\textbf{\Large Supplementary Material: Cavity Induced Topology in Graphene}
\end{center}
\hspace{5mm}
\begin{center}
{\large Ceren~B.~Dag and Vasil~Rokaj}
\end{center}

\vspace{5mm}

\section{\label{sec:TRS}Time Reversal Symmetry in Electron-Photon Systems}

Time-reversal symmetry (TRS) plays a key role in understanding the topological properties of condensed matter systems, as for example in Chern insulators~\cite{HaldaneChern} or Floquet engineering~\cite{Oka}. Here, we will discuss how TRS can be understood and described in systems where charged particles are coupled to the quantized photon fields. For this investigation we will rely on the Pauli-Fierz Hamiltonian describing non-relativistic electrons coupled to photons, also known as the minimal coupling Hamiltonian~\cite{spohn2004, cohen1997photons}. For a single electron in a periodic crystal potential coupled to light we have,
\begin{equation}\label{minimal coupling}
    \mathcal{H}=\frac{1}{2m}\left(\textrm{i}\hbar\nabla+ e\mathbf{A}\right)^2+V_{\textrm{crys}}(\mathbf{r})+\sum_{\lambda=x,y}\hbar\omega_c\left(b^{\dagger}_{\lambda}b_{\lambda}+\frac{1}{2}\right).
\end{equation}
In the above Hamiltonian for simplicity we assumed a single-mode photon field with frequency $\omega_c=c|\kappa_z|$ where $\kappa_z$ is photon momentum chosen along the $z$ direction. Under this choice the polarizations of the photon field are in the $(x,y)$ plane. Moreover, the operators $b^{\dagger}_{\lambda}, b_{\lambda}$ are the creation and annihilation operators of the photon field satisfying bosonic commutation relations $[b_{\lambda},b^{\dagger}_{\lambda^{\prime}}]=\delta_{\lambda\lambda^{\prime}}$. The effective mode volume is $\mathcal{V}$, $\epsilon_0$ is the vacuum permittivity and $m$ is the effective electron mass. For two-dimensional materials whose thickness is at the order of a single or a few atoms~\cite{Mak2022, Moiremarvels}, the variation of the photon field in the $z$ direction can be ignored. This is the well-known long-wavelength (or dipole) approximation, and the photon field takes the simple form~\cite{rokaj2017}
\begin{equation}
    \mathbf{A}=\sqrt{\frac{\hbar}{\epsilon_0\mathcal{V}2\omega_c}}\sum_{\lambda=x,y}\mathbf{e}_{\lambda}\left(b^{\dagger}_{\lambda}+b_{\lambda}\right) \label{eq:vectorPotLinPol}
\end{equation}
In the above expression we have chosen linearly polarized light with polarization vectors $\mathbf{e}_x$ and $\mathbf{e}_y$ as a starting point. Later we discuss the case of circularly polarized photons.

\subsection{TRS for Linearly Polarized Photons}

In classical physics, the momentum $\mathbf{p}$ of a particle and the classical vector potential $\mathbf{A}_{\rm{cl}}(t)$ responsible for a classical electric field $\mathbf{E}_{\rm{cl}}(t)$ transform under time-reversal $\mathcal{T}$ transform as~\cite{TRSb}
\begin{equation}
    \mathcal{T}(\mathbf{p})=-\mathbf{p}\;\; \textrm{and}\;\; \mathcal{T}(\mathbf{A}_{\rm{cl}})=-\mathbf{A}_{\rm{cl}}. \label{eq:TRStransformation}
\end{equation}
Eq.~\eqref{eq:TRStransformation} guarantees that the kinetic energy of the particle and the electric field $\mathbf{E}_{\rm{cl}}$ are invariant under TRS. The momentum operator in quantum mechanics transforms under TRS in the same way as the classical momentum $\mathcal{T}(-\textrm{i}\hbar \nabla)=\textrm{i}\hbar \nabla$. Again in classical physics, a linearly polarized electric field preserves TRS. These transformation rules must be preserved under quantization. Thus, for the minimal coupling Hamiltonian to be invariant under TRS, the quantized vector potential for linearly polarized photons must transform as, $\mathcal{T}(\mathbf{A})=-\mathbf{A}$. For this transformation to hold, the annihilation and creation photon operators must transform under $\mathcal{T}$ as follows 
\begin{equation}
    \mathcal{T}(b_{\lambda})=-b_{\lambda} \;\; \textrm{and}\;\; \mathcal{T}(b^{\dagger}_{\lambda})=-b^{\dagger}_{\lambda}.
\end{equation}
With the use of the above relations we find that the energy of the photon number operator $b^{\dagger}_{\lambda}b_{\lambda}$ is invariant under TRS $\mathcal{T}(b^{\dagger}_{\lambda}b_{\lambda})=b^{\dagger}_{\lambda}b_{\lambda}$. By combining all the transformation rules, we find that the minimal coupling Hamiltonian for linearly polarized light is invariant under TRS $\mathcal{T}(\mathcal{H})=\mathcal{H}$. We numerically confirmed this fact by considering graphene coupled to linearly polarized photons, and we found that no gap opens at Dirac nodes. This holds both with one, Fig.~\ref{Fig5}(c), and two linear polarizations, Fig.~\ref{SFig1}(f).

\subsection{TRS for Circularly Polarized Photons}
The linearly polarized photon field including both polarizations $\mathbf{e}_x$ and $\mathbf{e}_y$ can be equivalently written in terms of left $\mathbf{e}_L=(1,-\textrm{i})/\sqrt{2}$ and right $\mathbf{e}_R=(1,\textrm{i})/\sqrt{2}$ circular polarizations through the expressions $\mathbf{e}_x=(\mathbf{e}_R+\mathbf{e}_L)/\sqrt{2}$ and $\mathbf{e}_y=-\textrm{i}(\mathbf{e}_R-\mathbf{e}_L)/\sqrt{2}$~\cite{BaymQM}. Then the photon field takes the form
\begin{eqnarray}
 \mathbf{A}=\sqrt{\frac{\hbar}{\epsilon_0\mathcal{V}2\omega_c}}\left[\mathbf{e}_R\frac{1}{\sqrt{2}}\left(b_x-\textrm{i}b_y\right)+\mathbf{e}_R\frac{1}{\sqrt{2}}\left(b^{\dagger}_x-\textrm{i}b^{\dagger}_y\right)+\mathbf{e}_L\frac{1}{\sqrt{2}}\left(b_x+\textrm{i}b_y\right)+\mathbf{e}_L\frac{1}{\sqrt{2}}\left(b^{\dagger}_x+\textrm{i}b^{\dagger}_y\right)\right].
\end{eqnarray}
We can define the following set of annihilation and creation photon operators for left and right circularly polarized photons
\begin{eqnarray}
 &&b_L=\frac{1}{\sqrt{2}}\left(b_x-\textrm{i}b_y\right) \;\; \textrm{and}\;\; b^{\dagger}_L=\frac{1}{\sqrt{2}}\left(b^{\dagger}_x+\textrm{i}b^{\dagger}_y\right)\\
 &&b_R=\frac{1}{\sqrt{2}}\left(b_x+\textrm{i}b_y\right) \;\; \textrm{and}\;\;b^{\dagger}_R=\frac{1}{\sqrt{2}}\left(b^{\dagger}_x-\textrm{i}b^{\dagger}_y\right). \notag
\end{eqnarray}
We note that the left and right handed photon operators satisfy standard bosonic commutation relations $[b_L,b^{\dagger}_L]=[b_R,b^{\dagger}_R]=1$ and are independent $[b_L,b^{\dagger}_R]=0$. Thus, the photon field in terms of the left and right handed photons takes the form
\begin{eqnarray}
 \mathbf{A}=\sqrt{\frac{\hbar}{\epsilon_0\mathcal{V}2\omega_c}}\left[\mathbf{e}_Rb_L+\mathbf{e}_Rb^{\dagger}_R+\mathbf{e}_Lb_R+\mathbf{e}_Lb^{\dagger}_L\right].
\end{eqnarray}
Using the definition of the left and right polarized photon operators we find their transformation under time-reversal,
\begin{equation}
    \mathcal{T}(b_{L,R})=-b_{R,L}\;\; \textrm{and}\;\; \mathcal{T}(b^{\dagger}_{L,R})=-b^{\dagger}_{R,L}.
\end{equation}
Thus, we see that time-reversal exchanges left and right photon operators (up to a minus). The same also holds for left and right polarization vectors due to the imaginary unit, $\mathcal{T}(\mathbf{e}_R)=\mathbf{e}_L$ and $\mathcal{T}(\mathbf{e}_L)=\mathbf{e}_R$. Using all the transformation rules we find that photon field written in terms of the left and right circular polarization, transforms in the same way as the photon field written in terms of the linear polarizations, i.e., $\mathcal{T}(\mathbf{A})=-\mathbf{A}$. The energy of the photon field in terms of left and right circularly polarized operators takes the standard form 
$$\sum_{\lambda=x,y}\hbar \omega_c\left( b^{\dagger}_{\lambda}b_{\lambda}+\frac{1}{2}\right)=\sum_{\lambda=L,R}\hbar \omega_c \left(b^{\dagger}_{\lambda}b_{\lambda}+\frac{1}{2}\right).$$
Hence it is evident that the energy of the photon field, including both polarizations, is invariant under time-reversal. Thus, as long as we keep both polarizations for the mode the minimal coupling Hamiltonian preserves TRS, as expected. This is also numerically confirmed below in Fig.~\ref{SFig1}(f) by considering the graphene coupled to both left and right polarizations if their couplings and frequencies are exactly the same.   

\subsection{TRS Breaking}
However, if we eliminate either the left or the right circularly polarized photons, the TRS is broken. To show this we consider the case where we have only the left polarized photons and the right ones are completely eliminated. In this case the photon field has the form
\begin{equation}
    \mathbf{A}_L=\sqrt{\frac{\hbar}{\epsilon_0\mathcal{V}2\omega_c}}\left[\mathbf{e}_Rb_L+\mathbf{e}_Lb^{\dagger}_L\right].
\end{equation}
We apply the TRS operator on the left circularly polarized photon field, and find that
\begin{equation}
    \mathcal{T}(\mathbf{A}_L)=-\sqrt{\frac{\hbar}{\epsilon_0\mathcal{V}2\omega_c}}\left[\mathbf{e}_Lb_R+\mathbf{e}_Rb^{\dagger}_R\right]=-\mathbf{A}_R.
\end{equation}
This means that the left circularly polarized photon field is mapped to $-\mathbf{A}_R$, and thus TRS is broken. This is an extreme case where TRS is broken because the right circularly polarized photons are completely eliminated. However, TRS breaking also occurs if we have a field where both $\mathbf{A}_L$ and $\mathbf{A}_R$ are taken into account with different field strengths, $\mathbf{A}^{\prime}=\alpha_L \mathbf{A}_L +\alpha_R\mathbf{A}_R$ where $\alpha_R \neq \alpha_L$. Then one sees that this photon field does not satisfy the necessary condition for the TRS to be preserved, i.e.,~$\mathcal{T}\left(\mathbf{A}^{\prime}\right)\neq -\mathbf{A}^{\prime}$. This is the scenario which we investigate in the main text, and find that for graphene coupled to such a photon field a topological gap occurs at the Dirac node signaling the TRS breaking.   

\section{Tight-Binding Model for Graphene Coupled to Photons}

The aim of this section is to derive the tight-binding Hamiltonian for graphene coupled to photons with both polarizations starting from the minimal coupling Hamiltonian. 
Expanding the covariant kinetic energy in the minimal-coupling Hamiltonian in Eq.~(\ref{minimal coupling}) we have
\begin{equation}
    \mathcal{H}=\underbrace{-\frac{\hbar^2}{2m}\nabla^2+V_{\textrm{crys}}(\mathbf{r})}_{\textrm{Matter:}\;\; \mathcal{H}_{\textrm{m}}
    }+\underbrace{\frac{\textrm{i}e\hbar}{m}\mathbf{A}\cdot\nabla}_{\textrm{Photon-Matter}: \;\; \mathcal{H}_{\textrm{pm}}} +\underbrace{\frac{e^2}{2m}\hat{\mathbf{A}}^2 +\sum_{\lambda=x,y}\hbar\omega_c\left(b^{\dagger}_{\lambda}b_{\lambda}+\frac{1}{2}\right)}_{\textrm{Photonic}:\;\; \mathcal{H}_p} .
    \end{equation}
where the external potential is periodic under Bravais lattice translations $V_{\textrm{crys}}(\mathbf{r}+\mathbf{R}_\mathbf{j})=V_{\textrm{crys}}(\mathbf{r})$ with $\mathbf{R}_{\mathbf{j}}$ being the Bravais lattice vectors~\cite{Mermin}. The single-mode photon field in the long-wavelength (homogeneous) limit is Eq.~\eqref{eq:vectorPotLinPol}
with the index $\lambda$ indicates the two orthogonal linear polarizations $\mathbf{e}_x$ and $\mathbf{e}_y$. In the Hamiltonian we have a purely photonic part $\mathcal{H}_p$ which depends only on the annihilation $b_{\lambda}$ and creation  $ b^{\dagger}_{\lambda}$ operators of the photon field. Substituting the expression for the vector potential $\mathbf{A}$ and introducing the diamagnetic shift $\omega_D$
\begin{eqnarray}\label{plasma frequency}
    \omega_D=\sqrt{\frac{e^2}{m\epsilon_0 \mathcal{V}}} \notag
\end{eqnarray}
the photonic part $\mathcal{H}_p$ takes the form
\begin{eqnarray}\label{photonicpart}
\mathcal{H}_p=\sum^{2}_{\lambda=1}\left[\hbar\omega_c\left(b^{\dagger}_{\lambda}b_{\lambda}+\frac{1}{2}\right)+\frac{\hbar \omega^2_D}{4\omega_c}\left(b_{\lambda}+b^{\dagger}_{\lambda}\right)^2\right].
\end{eqnarray}
The photonic part $\mathcal{H}_p$ can be brought into diagonal form by introducing a new set of bosonic operators: $a^{\dagger}_{\lambda}$ and $a_{\lambda}$
\begin{eqnarray}\label{boperators}
a_{\lambda}=\frac{1}{2\sqrt{\omega_c\omega}}\left[b_{\lambda}\left(\omega+\omega_c\right)+b^{\dagger}_{\lambda}\left(\omega-\omega_c\right)\right] \;\; \textrm{and}\;\; a^{\dagger}_{\lambda}=\frac{1}{2\sqrt{\omega_c \omega}} \left[b_{\lambda}\left(\omega-\omega_c\right)+b^{\dagger}_{\lambda}\left(\omega+\omega_c\right)\right], \;\; \textrm{with}\;\; \omega=\sqrt{\omega^2_c+\omega^2_D}.\nonumber\\
\end{eqnarray}
The frequency $\omega$ is the dressed cavity frequency which depends on the bare photon/cavity frequency $\omega_c$ and the diamagnetic shift $\omega_D$~\cite{Rokaj2deg}. The operators $a_{\lambda},a^{\dagger}_{\lambda}$ satisfy bosonic commutation relations $[a_{\lambda},a^{\dagger}_{\lambda^{\prime}}]=\delta_{\lambda,\lambda^{\prime}}$ for $\lambda,\lambda^{\prime}=1,2$. $\mathcal{H}_p$ is equal to the sum of two non-interacting harmonic oscillators in terms of this new set of operators,  
\begin{equation}\label{Hpinb}
\mathcal{H}_{p}=\sum^{2}_{\lambda=1}\hbar\omega\left(a^{\dagger}_{\lambda}a_{\lambda}+\frac{1}{2}\right)
\end{equation}
and the quantized vector potential $\mathbf{A}$ is
\begin{eqnarray}\label{Ainb}
\mathbf{A}=\left(\frac{\hbar}{\epsilon_0\mathcal{V}}\right)^{\frac{1}{2}}\sum^{2}_{\lambda=1}\frac{\mathbf{e}_{\lambda}}{\sqrt{2\omega}}\left(a_{\lambda}+a^{\dagger}_{\lambda}\right).
\end{eqnarray}
Substituting back into $\mathcal{H}$ the expression for the photonic part $\mathcal{H}_p$ we have
\begin{equation}\label{no A2}
    \mathcal{H}=\underbrace{-\frac{\hbar^2}{2m}\nabla^2+V_{\textrm{crys}}(\mathbf{r})}_{\textrm{Matter:} \mathcal{H}_m}+\underbrace{\frac{\textrm{i}e\hbar}{m}\mathbf{A}\cdot\nabla}_{\textrm{Photon-Matter:} \mathcal{H}_{pm}} +\underbrace{\sum^{2}_{\lambda=1}\hbar\omega\left(a^{\dagger}_{\lambda}a_{\lambda}+\frac{1}{2}\right)}_{\textrm{Photon:} \mathcal{H}_p} .
    \end{equation}
with the index $\lambda$ indicating the two orthogonal linear polarizations $\mathbf{e}_x$ and $\mathbf{e}_y$. Graphene consists of two sublattices, $A$ and $B$, and as a consequence the tight-binding ansatz wavefunction consists of two components, one for each sublattice~\cite{Bena_graphene},
\begin{equation}\label{TBansatz}
    \Psi_{\mathbf{k}}(\mathbf{r})=a_{\mathbf{k}}\Psi^{A}_{\mathbf{k}}(\mathbf{r})+b_{\mathbf{k}}\Psi^{B}_{\mathbf{k}}(\mathbf{r})=\sum_{\mathbf{j}}e^{\textrm{i}\mathbf{k}\cdot \mathbf{R}_{\mathbf{j}}}\left[a_{\mathbf{k}}\phi_A(\mathbf{r}-\mathbf{R}_{\mathbf{j}})+b_{\mathbf{k}}\phi_B(\mathbf{r}-\mathbf{R}^B_{\mathbf{j}})\right],
\end{equation}
where $\mathbf{R}_{\mathbf{j}}=j_1\mathbf{a}_1+j_2\mathbf{a}_2$ are the Bravais vectors of sublattice $A$ with $\mathbf{a}_1=\mathbf{e}_x\alpha\sqrt{3}/2+\mathbf{e}_y3\alpha/2$ and $\mathbf{a}_2=-\mathbf{e}_x\alpha\sqrt{3}/2+\mathbf{e}_y3\alpha/2$. The Bravais vectors for the sublattice $B$ are $\mathbf{R}^B_{\mathbf{j}}=\mathbf{R}_{\mathbf{j}}+\bm{\delta}_3$ with $\bm{\delta}_3=-\alpha\mathbf{e}_y$. To derive the tight-binding Hamiltonian for graphene coupled to photons we apply $\mathcal{H}$ on the tight-binding wavefunction ansatz for graphene,
\begin{eqnarray}
 \mathcal{H}_{\mathbf{k}}= \left(\begin{tabular}{ c c }
		$\langle \Psi^A_{\mathbf{k}}|\mathcal{H}|\Psi^A_{\mathbf{k}}\rangle$ &$ \langle \Psi^A_{\mathbf{k}}|\mathcal{H}|\Psi^B_{\mathbf{k}}\rangle$ \\
	    $\langle \Psi^A_{\mathbf{k}}|\mathcal{H}|\Psi^B_{\mathbf{k}}\rangle^*$ &$ \langle \Psi^B_{\mathbf{k}}|\mathcal{H}|\Psi^B_{\mathbf{k}}\rangle $ 
	\end{tabular}\right).
\end{eqnarray}
The minimal coupling Hamiltonian consists of the matter Hamiltonian $\mathcal{H}_m$, the light-matter part $\mathcal{H}_{\rm{pm}}$, and the purely photonic part $\mathcal{H}_p$ which acts trivially to the tight-binding wavefunction. Within a tight-binding model, a solid is viewed as a collection of atoms with electrons well localized around the atoms. Thus, it is convenient to write the matter Hamiltonian of the crystal $\mathcal{H}_{\textrm{m}}$ as a sum of the Hamiltonian describing an atom $\mathcal{H}_{\textrm{at}}$ and the potential $\delta V(\mathbf{r})$ which describes the rest of the crystal, $ \mathcal{H}_{\textrm{m}}=\mathcal{H}_{\textrm{at}}+\delta V(\mathbf{r})$. We note that the atom potential $V_{\textrm{at}}$ together with $\delta V(\mathbf{r})$ gives the crystal potential, $V_{\textrm{crys}}(\mathbf{r})=V_{\textrm{at}}(\mathbf{r})+\delta V(\mathbf{r})$. It is important to mention that for the construction of the tight-binding ansatz in (\ref{TBansatz}) the localized states of $\mathcal{H}_{\textrm{at}}$ are used~\cite{Mermin, Bena_graphene}. We now project the full minimal coupling Hamiltonian on the tight-binding ansatz wavefunction. For the matter Hamiltonian $\mathcal{H}_m$ we have
\begin{eqnarray}
 \mathcal{H}^m_{\mathbf{k}}= \left(\begin{tabular}{ c c }
		$\langle \Psi^A_{\mathbf{k}}|\mathcal{H}_m|\Psi^A_{\mathbf{k}}\rangle$ &$ \langle \Psi^A_{\mathbf{k}}|\mathcal{H}_m|\Psi^B_{\mathbf{k}}\rangle$ \\
	    $\langle \Psi^A_{\mathbf{k}}|\mathcal{H}_m|\Psi^B_{\mathbf{k}}\rangle^*$ &$ \langle \Psi^B_{\mathbf{k}}|\mathcal{H}_m|\Psi^B_{\mathbf{k}}\rangle $ 
	\end{tabular}\right) 
\end{eqnarray}
The diagonal elements $ \langle \Psi^B_{\mathbf{k}}|\mathcal{H}_m|\Psi^B_{\mathbf{k}}\rangle $ and $ \langle \Psi^A_{\mathbf{k}}|\mathcal{H}_m|\Psi^A_{\mathbf{k}}\rangle $  result in the terms which are beyond the nearest neighbors, and hence we eliminate them. Thus, we only compute the off-diagonals,
\begin{equation}
  \langle \Psi^A_{\mathbf{k}}|\mathcal{H}_m|\Psi^B_{\mathbf{k}}\rangle=  \langle \Psi^A_{\mathbf{k}}|\mathcal{H}_{at}|\Psi^B_{\mathbf{k}}\rangle +  \langle \Psi^A_{\mathbf{k}}|\delta V|\Psi^B_{\mathbf{k}}\rangle= E_A \underbrace{\langle \Psi^A_{\mathbf{k}}|\Psi^B_{\mathbf{k}}\rangle}_{=0}+\sum_{\mathbf{j},\mathbf{q}} e^{\textrm{i}\mathbf{k}\cdot(\mathbf{R}_{\mathbf{j}}-\mathbf{R}_{\mathbf{q}})} \int d^3r \phi^*_A(\mathbf{r}-\mathbf{R}_{\mathbf{q}})\delta V(\mathbf{r}) \phi_B(\mathbf{r}-\mathbf{R}_{\mathbf{j}}-\bm{\delta}_3).
\end{equation}
Next we perform the coordinate shift $\mathbf{r}\rightarrow \mathbf{r} +\mathbf{R}_{\mathbf{q}}$, define $\mathbf{R}_{\mathbf{f}}=\mathbf{R}_{\mathbf{j}}-\mathbf{R}_{\mathbf{q}}$ and have
\begin{equation}
 \langle \Psi^A_{\mathbf{k}}|\mathcal{H}_m|\Psi^B_{\mathbf{k}}\rangle=\sum_{\mathbf{f}}e^{\textrm{i}\mathbf{k}\cdot \mathbf{R}_{\mathbf{f}}} \int d^3r \phi^*_A(\mathbf{r})\delta V(\mathbf{r}) \phi_B(\mathbf{r}-\mathbf{R}_{\mathbf{f}}-\bm{\delta}_3)  = \sum_{\mathbf{f}}e^{\textrm{i}\mathbf{k}\cdot \mathbf{R}_{\mathbf{f}}} t(|\mathbf{R}_{\mathbf{f}}+\bm{\delta}_3|)
\end{equation}
where $t(|\mathbf{R}_{\mathbf{f}}+\bm{\delta}_3|)$ is the tunneling matrix element due to the potential $\delta V(\mathbf{r})$, which depends only on the distance between different lattice points. We take into account only the nearest neighbor tunneling with the vectors $\mathbf{f}=(f_1,f_2)=(0,0)$, $\mathbf{f}=(1,0)$ and $\mathbf{f}=(0,1)$ which have the same distance from the origin
\begin{eqnarray}
 |\mathbf{R}_{0,0}+\bm{\delta}_3|=|(0,-\alpha)|=\alpha,\;\; |\mathbf{R}_{1,0}+\bm{\delta}_3|=|(\alpha\sqrt{3}/2,\alpha/2)|=\alpha,\;\;|\mathbf{R}_{0,1}+\bm{\delta}_3|=|(-\alpha\sqrt{3}/2,\alpha/2)|=\alpha.
\end{eqnarray}
This leads to tunneling elements of precisely the same strength $t(|\mathbf{R}_{0,0}+\bm{\delta}_3|)=t(|\mathbf{R}_{1,0}+\bm{\delta}_3|)=t(|\mathbf{R}_{0,1}+\bm{\delta}_3|)\equiv t$ and we find
\begin{equation}
    \langle \Psi^A_{\mathbf{k}}|\mathcal{H}_m|\Psi^B_{\mathbf{k}}\rangle= t \left[1+e^{\textrm{i}\mathbf{k}\cdot\mathbf{a}_1} + e^{\textrm{i}\mathbf{k}\cdot\mathbf{a}_2}\right] = t \hspace{1mm} h(\mathbf{k}).
\end{equation}
In the last step we also assumed that the tunneling elements for the nearest neighboring bonds are equal to $t$. Thus, for the matter Hamiltonian we obtain 
\begin{eqnarray}
 \mathcal{H}^m_{\mathbf{k}}=t \left(\begin{tabular}{ c c }
		$0$ &$ h(\mathbf{k})$ \\
	    $h^*(\mathbf{k})$ &$ 0 $ 
	\end{tabular}\right) \;\; \textrm{where}\;\; h(\mathbf{k})=\left[1+e^{\textrm{i}\mathbf{k}\cdot\mathbf{a}_1} + e^{\textrm{i}\mathbf{k}\cdot\mathbf{a}_2}\right]. 
\end{eqnarray}
The Hamiltonian $\mathcal{H}^m_{\mathbf{k}}$ describes the well-known two-band model of graphene~\cite{Bena_graphene,bernevig2013topological}. In terms of Pauli matrices $\sigma_{1,2}$ the matter Hamiltonian reads
\begin{eqnarray}
 \mathcal{H}^m_{\mathbf{k}} &=& d_1(\mathbf{k}) \sigma_1 + d_2(\mathbf{k}) \sigma_2,\;\; \textrm{where}\;\; d_1(\mathbf{k}) = t \left[ \cos(\mathbf{k}\cdot \mathbf{a}_1)+\cos(\mathbf{k}\cdot \mathbf{a}_2) + 1 \right], \;\;
 d_2(\mathbf{k}) = t \left[ \sin(\mathbf{k}\cdot \mathbf{a}_1)+\sin(\mathbf{k}\cdot \mathbf{a}_2)\right].\nonumber\\
\end{eqnarray}
Now we apply the photon-matter part $\mathcal{H}_{\textrm{pm}}$ to the tight-binding wavefunction 
\begin{eqnarray}
 \mathcal{H}^{\textrm{pm}}_{\mathbf{k}}= \left(\begin{tabular}{ c c }
		$\langle \Psi^A_{\mathbf{k}}|\mathcal{H}_{\textrm{pm}}|\Psi^A_{\mathbf{k}}\rangle$ &$ \langle \Psi^A_{\mathbf{k}}|\mathcal{H}_{\textrm{pm}}|\Psi^B_{\mathbf{k}}\rangle$ \\
	    $\langle \Psi^A_{\mathbf{k}}|\mathcal{H}_{\textrm{pm}}|\Psi^B_{\mathbf{k}}\rangle^{*}$ &$ \langle \Psi^B_{\mathbf{k}}|\mathcal{H}_{\textrm{pm}}|\Psi^B_{\mathbf{k}}\rangle $ 
	\end{tabular}\right)  \label{PMMatrix}
\end{eqnarray}
As before, we neglect the diagonal terms which result in the tunneling beyond the nearest neighbor bonds. Thus, we only need to compute $\langle \Psi^A_{\mathbf{k}}|\nabla|\Psi^B_{\mathbf{k}}\rangle$ which after performing the transformations $\mathbf{r}\rightarrow \mathbf{r} +\mathbf{R}_{\mathbf{q}}$ and $\mathbf{R}_{\mathbf{f}}=\mathbf{R}_{\mathbf{j}}-\mathbf{R}_{\mathbf{q}}$ is
\begin{equation}
    \textrm{i}\hbar\langle \Psi^A_{\mathbf{k}}|\nabla|\Psi^B_{\mathbf{k}}\rangle=\textrm{i}\hbar\sum_{\mathbf{f}}e^{\textrm{i}\mathbf{k}\cdot \mathbf{R}_{\mathbf{f}}} \int d^2r \phi^*_A(\mathbf{r})\nabla \phi_B(\mathbf{r}-\mathbf{R}_{\mathbf{f}}-\bm{\delta}_3) \label{eq:S27}
\end{equation}
Here we are interested in the nearest neighbor tunneling which implies that the Bravais points of interest $\mathbf{R}_{\mathbf{f}}$  are small. Thus we can Taylor-expand $\phi_B(\mathbf{r}-\mathbf{R}_{\mathbf{f}}-\bm{\delta}_3)$ and keep only up to the first order in the series $\phi_B(\mathbf{r}-\mathbf{R}_{\mathbf{f}}-\bm{\delta}_3)= \phi_B(\mathbf{r})-(\mathbf{R}_{\mathbf{f}}+\bm{\delta}_3)\cdot \nabla \phi_B(\mathbf{r})$. Substituting the latter in Eq.~\eqref{eq:S27}, we find
\begin{eqnarray}
    &&\textrm{i}\hbar\langle \Psi^A_{\mathbf{k}}|\nabla|\Psi^B_{\mathbf{k}}\rangle=\\
    &&\sum_{\mathbf{f}}e^{\textrm{i}\mathbf{k}\cdot \mathbf{R}_{\mathbf{f}}} \textrm{i}\hbar \int d^2r \phi^*_A(\mathbf{r})\left[\nabla \phi_B(\mathbf{r})-\mathbf{e}_x R^x_{\mathbf{f}}\partial^2_x \phi_B(\mathbf{r})-\mathbf{e}_y (R^y_{\mathbf{f}}-\alpha)\partial^2_y \phi_B(\mathbf{r}) -\mathbf{e}_yR^x_{\mathbf{f}}\partial_x\partial_y\phi_B(\mathbf{r})-\mathbf{e}_x(R^y_{\mathbf{f}}-\alpha)\partial_x\partial_y\phi_B(\mathbf{r})\right]\nonumber
\end{eqnarray}
The atomic wavefunctions $\phi_A(\mathbf{r}), \phi_B(\mathbf{r})$ are either odd or even with respect to parity due to the symmetries of the atomic potential. The first-order derivatives will change the parity of the wavefunction and integrating over symmetric boundary leads to zero. Thus, the only the quadratic terms $\partial^2_x$ and $\partial^2_y$ give a non-zero contribution.
\begin{eqnarray}
 \textrm{i}\hbar\langle \Psi^A_{\mathbf{k}}|\nabla|\Psi^B_{\mathbf{k}}\rangle&=&\sum_{\mathbf{f}}e^{\textrm{i}\mathbf{k}\cdot \mathbf{R}_{\mathbf{f}}} \textrm{i}\hbar\int d^2r \phi^*_A(\mathbf{r})\left[-\mathbf{e}_x R^x_{\mathbf{f}}\partial^2_x \phi_B(\mathbf{r})-\mathbf{e}_y (R^y_{\mathbf{f}}-\alpha)\partial^2_y \phi_B(\mathbf{r})\right]\nonumber\\
 &=&-\textrm{i}\hbar\sum_{\mathbf{f}}e^{\textrm{i}\mathbf{k}\cdot \mathbf{R}_{\mathbf{f}}} \left[\mathbf{e}_xR^x_{\mathbf{f}}t_x+\mathbf{e}_y(R^y_{\mathbf{f}}-\alpha)t_y\right],
\end{eqnarray}
where $t_x$ and $t_y$ are the results of integration of the integrals over $\mathbf{r}$. Since we consider only the nearest neighbor bonds which are the points $\mathbf{f}=(0,0), (1,0)$ and $(0,1)$, we can explicitly write
\begin{eqnarray}
 \textrm{i}\hbar\langle \Psi^A_{\mathbf{k}}|\nabla|\Psi^B_{\mathbf{k}}\rangle&=&-\textrm{i}\hbar \Big[e^{\textrm{i}\mathbf{k}\cdot \mathbf{R}_{00}}\left(\mathbf{e}_x R^x_{00}t_x +\mathbf{e}_y(R^y_{00}-\alpha)t_y\right)+e^{\textrm{i}\mathbf{k}\cdot \mathbf{R}_{10}}\left(\mathbf{e}_x R^x_{10}t_x +\mathbf{e}_y(R^y_{10}-\alpha)t_y\right)\nonumber\\
 &+&e^{\textrm{i}\mathbf{k}\cdot \mathbf{R}_{01}}\left(\mathbf{e}_x R^x_{01}t_x +\mathbf{e}_y(R^y_{01}-\alpha)t_y\right)\Big].
\end{eqnarray}
Further we use the expressions for the $x$ and $y$ components of the Bravais lattice vectors $R^x_{\mathbf{f}}=\alpha\sqrt{3}(f_1-f_2)/2$ and $R^y_{\mathbf{f}}=3\alpha(f_1+f_2)/2$ for the nearest neighbor sites  $\mathbf{f}=(0,0), (1,0), (0,1)$, and we find
\begin{eqnarray}
  &&\textrm{i}\hbar\langle \Psi^A_{\mathbf{k}}|\nabla|\Psi^B_{\mathbf{k}}\rangle=\mathbf{e}_x \alpha t_x f_x(\mathbf{k})+\mathbf{e}_y \alpha t_yf_y(\mathbf{k})\;\;\; \textrm{where}\;\;\;\nonumber\\
  &&f_x(\mathbf{k})=-\textrm{i}\left(e^{\textrm{i}\mathbf{k}\cdot \mathbf{a}_1}-e^{\textrm{i}\mathbf{k}\cdot\mathbf{a}_2}\right) \frac{\sqrt{3}}{2}\;\; \textrm{and}\;\; f_y(\mathbf{k})=-\textrm{i}\left(-1+\frac{e^{\textrm{i}\mathbf{k}\cdot\mathbf{a}_1}}{2}+\frac{e^{\textrm{i}\mathbf{k}\cdot\mathbf{a}_2}}{2}\right).
\end{eqnarray}
Performing the same computation for the second off-diagonal term in $\mathcal{H}^{\rm{pm}}_{\mathbf{k}}$, we obtain
\begin{eqnarray}
\mathcal{H}^{\rm{pm}}_{\mathbf{k}}= \alpha\frac{e}{m}\mathbf{A}\cdot \left(\begin{tabular}{ c c }
		0 &$ \mathbf{e}_x t_x f_x(\mathbf{k})+\mathbf{e}_y  t_y f_y(\mathbf{k})$ \\
	    $\mathbf{e}_x t_x f^{*}_x(\mathbf{k})+\mathbf{e}_y t_y f^{*}_y(\mathbf{k})$ & $ 0 $
	\end{tabular}\right). \label{eq:S32}
\end{eqnarray}
It is important to note that we have not set a specific choice for the photon field polarization to obtain the expression above, and thus the result is general within the dipole approximation. We write Eq.~\eqref{eq:S32} in the spinor basis,
\begin{eqnarray}
\mathcal{H}^{\rm{pm}}_{\mathbf{k}}=\alpha \frac{e}{m}\mathbf{A}\cdot\left[\mathbf{P}_1(\mathbf{k})\sigma_1 + \mathbf{P}_2(\mathbf{k})\sigma_2\right],
\end{eqnarray}
where the $\mathbf{k}$-dependent functions are now vectors,
\begin{eqnarray}
  \mathbf{P}_1(\mathbf{k}) &=& -\mathbf{e}_x  \frac{\sqrt{3}}{2} t_x \left(\sin(\mathbf{k}\cdot \mathbf{a}_1)-\sin(\mathbf{k}\cdot \mathbf{a}_2) \right) -  \mathbf{e}_y \frac{1}{2}   t_y \left(\sin(\mathbf{k}\cdot \mathbf{a}_1)+\sin(\mathbf{k}\cdot \mathbf{a}_2) \right),\\ \mathbf{P}_2(\mathbf{k}) &=& \mathbf{e}_x  \frac{\sqrt{3}}{2}   t_x  \left(\cos(\mathbf{k}\cdot \mathbf{a}_1)-\cos(\mathbf{k}\cdot \mathbf{a}_2) \right) + \mathbf{e}_y   t_y \left(-1+\frac{1}{2} \cos(\mathbf{k}\cdot \mathbf{a}_1)+\frac{1}{2} \cos(\mathbf{k}\cdot \mathbf{a}_2) \right).\notag
\end{eqnarray}
We note that $t_x=t_y$ is assumed for the calculations which should hold for an unstrained graphene. By adding all the terms ---$\mathcal{H}^{\rm{m}}_{\mathbf{k}},\mathcal{H}^{\rm{pm}}_{\mathbf{k}}$ and $\mathcal{H}_p$---, we find the expression for the tight-binding model of graphene coupled to a cavity photon field,
\begin{eqnarray}
    \mathcal{H}_{\mathbf{k}}=d_1(\mathbf{k}) \sigma_1 + d_2(\mathbf{k}) \sigma_2+\alpha \frac{e}{m}\mathbf{A}\cdot\left[\mathbf{P}_1(\mathbf{k})\sigma_1 + \mathbf{P}_2(\mathbf{k})\sigma_2\right] + \sum_{\lambda=x,y}\hbar \omega\left(a^{\dagger}_{\lambda}a_{\lambda}+\frac{1}{2}\right)
\end{eqnarray}
Finally using the expression for the photon field in terms of the left and right handed photons 
\begin{eqnarray}
 \mathbf{A}=\underbrace{\sqrt{\frac{\hbar}{\epsilon_0\mathcal{V}2\omega}}}_{A_0}\left[\mathbf{e}_R a_L+\mathbf{e}_R a^{\dagger}_R+\mathbf{e}_L a_R+\mathbf{e}_L a^{\dagger}_L\right].
\end{eqnarray}
we obtain the corresponding expression for the two-polarization model
\begin{eqnarray}
    \mathcal{H}_{\mathbf{k}}=d_1(\mathbf{k}) \sigma_1 + d_2(\mathbf{k}) \sigma_2+\alpha\frac{e}{m}A_0\left[\mathbf{e}_R a_L+\mathbf{e}_R a^{\dagger}_R+\mathbf{e}_L a_R+\mathbf{e}_L a^{\dagger}_L\right]\cdot\left[\mathbf{P}_1(\mathbf{k})\sigma_1 + \mathbf{P}_2(\mathbf{k})\sigma_2\right] + \sum_{\lambda=R,L}\hbar \omega\left(a^{\dagger}_{\lambda}a_{\lambda}+\frac{1}{2}\right). \label{eq:minimalcouplingTBHam}\nonumber\\
\end{eqnarray}

\begin{figure}
\centering
\includegraphics[width=0.98 \columnwidth]{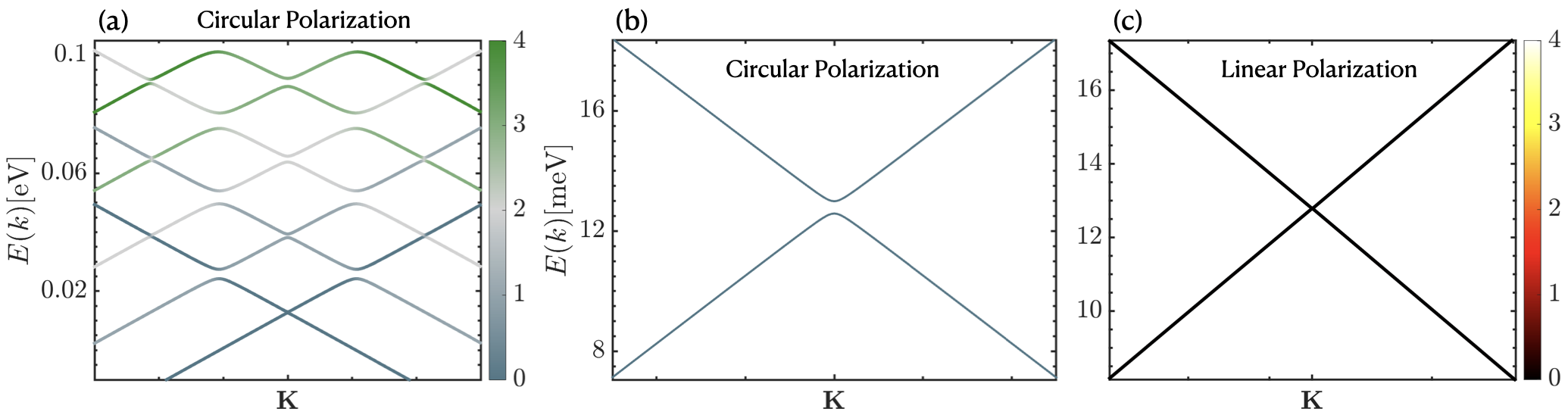}
\caption{(a) Tight-binding band structure around the Dirac node $\mathbf{K}$ calculated with exact diagonalization with a truncated photonic Hilbert space of maximum photon number $\langle a_{R}^{\dagger}a_{R}\rangle_{\textrm{max}}=4$. Frequency is set to be $\omega_c=6.28$ THz and $m=0.02m_e$. (b) Focus on the vacuum band gap. It is idential at the other Dirac node $\mathbf{K}'$. (c) Dirac nodes do not open when the cavity is linearly polarized in either direction because time-reversal symmetry is preserved, see Sec.~\ref{sec:TRS}. Rest of the parameters are the same with previous figures. }
\label{Fig5}
\end{figure}

\subsection{Photo-electron band structure}

The band structure of the tight-binding model around the Dirac nodes in the single-polarization limit is shown in Fig.~\ref{Fig5}(a) where the color code denotes the right-circularly polarized photon population in the photo-electron bands. The Dirac bands reported in the main text, Fig.~2, are almost identical to the band structure found by the tight-binding model. Let us note that one needs to tune the Fermi velocity $v_{\textrm{F}}$ to see this correspondence, which is the reason of fixing $v_{\textrm{F}}=0.21$ [a.u.]. Fig.~\ref{Fig5}(b) focuses on the vacuum bands in (a), which shows the topological band gap opening at Dirac nodes. The gap at the other Dirac node at $\mathbf{K}'$ is the same [not shown], and these vacuum bands are virtually the same with the Dirac vacuum bands. 

\begin{figure}
\centering
\includegraphics[width=0.95\columnwidth]{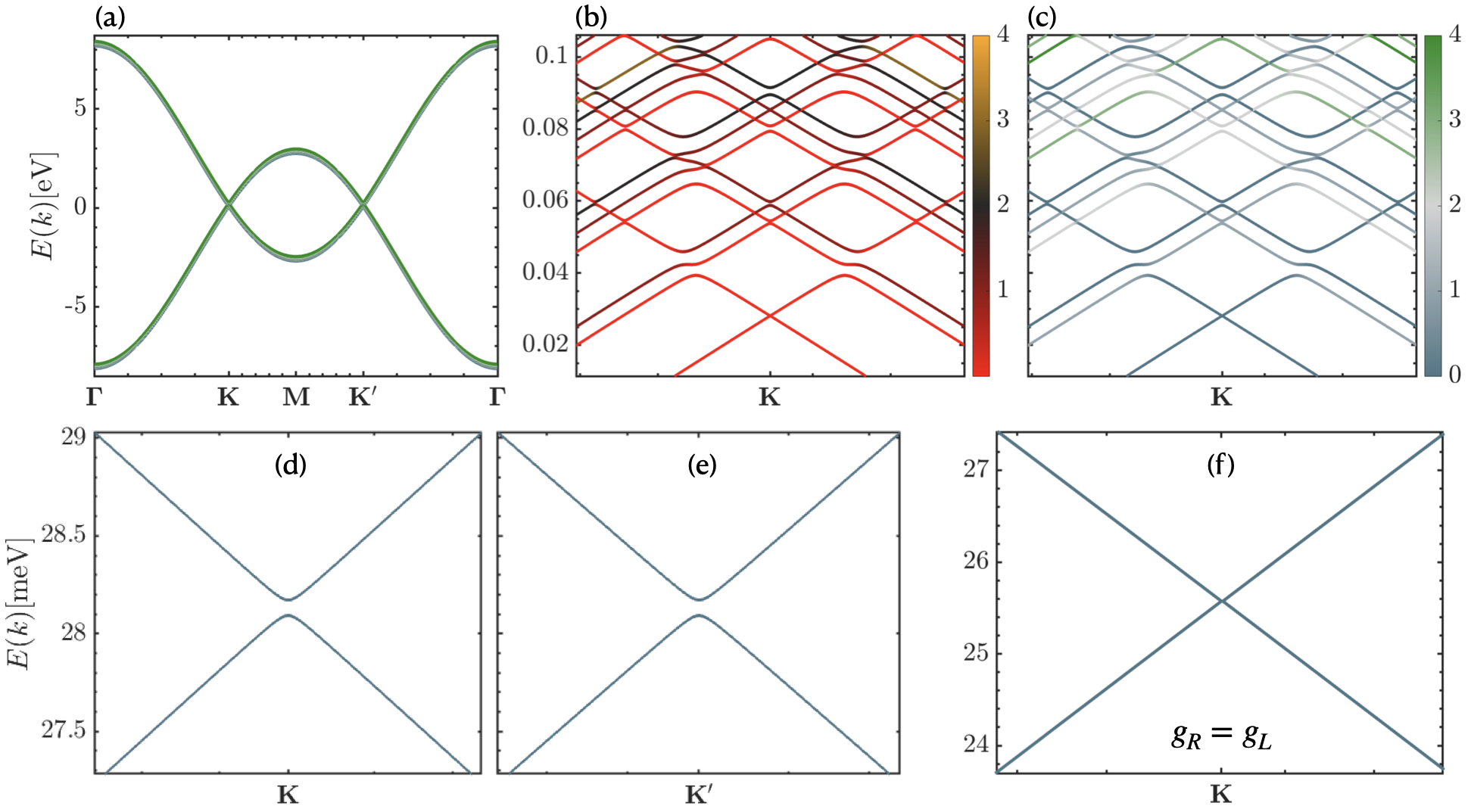}
\caption{Tight-binding model band structure results for two-polarization setup with two Faraday rotators and hence with an incurred frequency shift ratio $\gamma=1.2$ (see text) calculated with exact diagonalization with a truncated photonic Hilbert space of maximum photon number $\langle a_{\lambda}^{\dagger}a_{\lambda}\rangle_{\textrm{max}}=4$. The cavity frequency is set to be $\omega_c=6.28$ THz and $m=0.02m_e$. (a) Band structure for the entire Brillouin zone plotted at the high symmetry points of graphene. (b)-(c) Focused around the $\mathbf{K}$ point where the color code denotes (b) left-circularly and (c) right-circularly polarized photon populations. (d)-(e) Focus on the vacuum gap at both Dirac nodes. (f) Focus on the Dirac node when the time-reversal symmetry is preserved due to  $\omega_R=\omega_L$ and $g_R=g_L$ or equivalently when the cavity is polarized linearly in both directions. Dirac nodes are protected by the preserved TRS in this case.}
\label{SFig1}
\end{figure}

Fig.~\ref{SFig1} features the band structure of a two-polarization model where two Faraday rotators need to be used to create a frequency shift between left- and right-circularly polarized photons. By fixing $\gamma=1.2$ in $\omega_2 = \gamma \omega_1$ which defines $\Omega_{L(R)}=\sqrt{\omega_{1(2)}^2+\omega_D^2}$, we find that the Dirac bands are again almost identical to the band structure of the tight-binding model. Fig.~\ref{SFig1}(a) shows the overall band structure at the fundamental cavity frequency $\omega_1=6.28$ THz, whereas Fig.~\ref{SFig1}(b-c) focuses around a Dirac node. As seen in (d-e), tight-binding model predicts the vacuum band splitting at both Dirac nodes. Let us note that we focus on this particular two-polarization model to study the higher photo-electron bands, because this model with the chosen parameters leads to Dirac bands which are compatible with the tight-binding band structure. The Berry curvature results below will also make it clear how the band topology of the photo-electron bands can be captured by the Dirac bands in this particular model. 

In all figures plotted in this section, we use $\mathcal{V}_{L}=  \chi \left(2\pi c/ \omega_{c}\right)^3$ where $\chi=5\times10^{-4}$ which is in experimental interval \cite{paravicini2019magneto} and an effective mass of $m=0.02m_e$. 

\subsection{Berry curvature calculations}

In this section, we plot the Berry curvature for different photo-electron bands, and show how the Chern numbers directly follow from the Berry curvature, and Berry phase counting. This section also numerically proves that the Dirac model captures the band structure physics found by the tight-binding model for the parameters we used in this work. In all figures plotted in this section, we use $\mathcal{V}_{L}=  \chi \left(2\pi c/ \omega_{c}\right)^3$ where $\chi=5\times10^{-4}$ which is in experimental interval \cite{paravicini2019magneto} and an effective mass of $m=0.02m_e$. 

\begin{figure}
\centering
\includegraphics[width=1 \columnwidth]{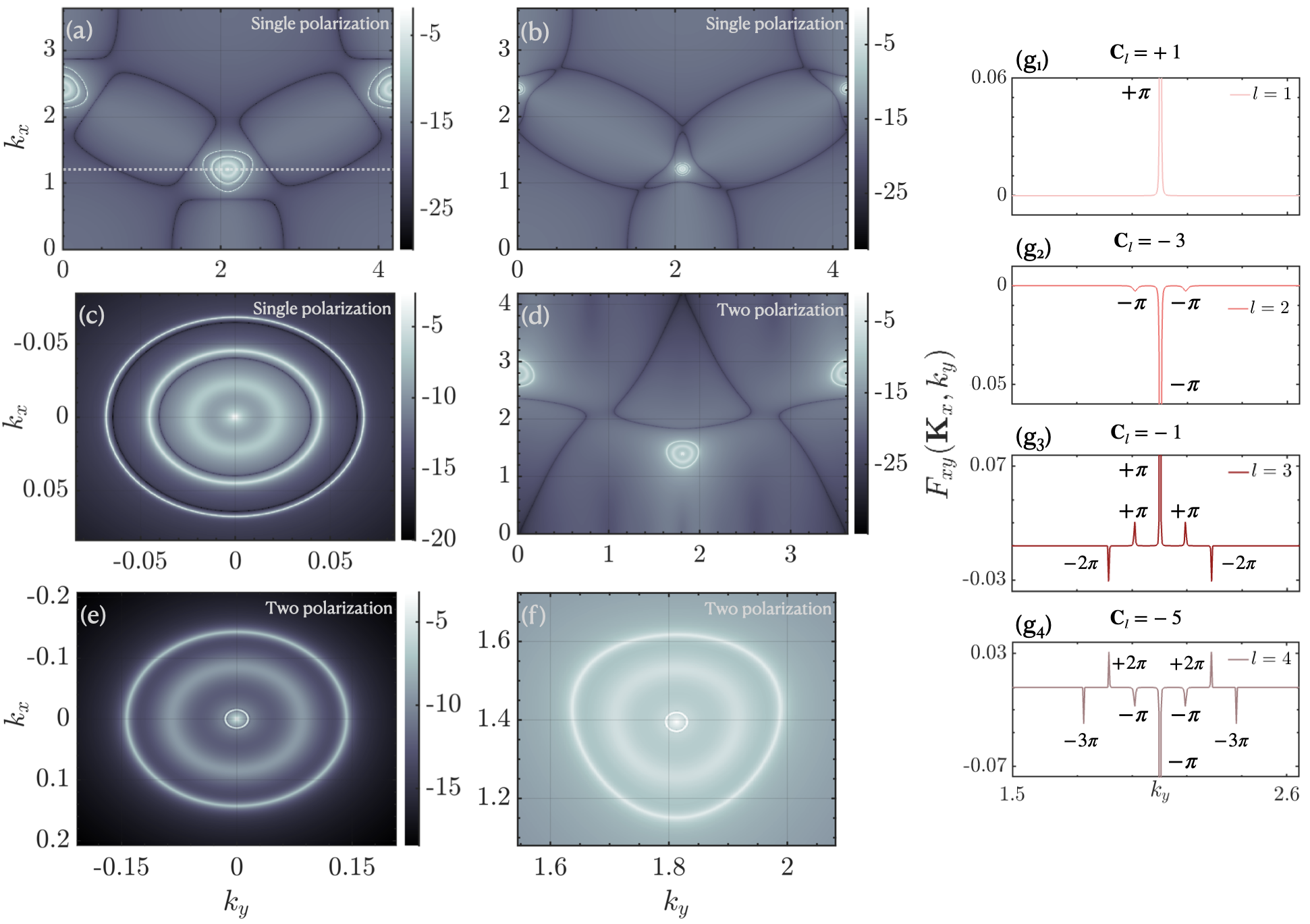}
\caption{(a-f) The magnitude of the Berry curvature for $l=4$, $\log|F_{l,xy}(k_x,k_y)|$ for single and two polarization models. (a) Berry curvature magnitude calculated with the tight-binding model in the single-polarization limit with a frequency of $\omega_c=200$ THz to be able to resolve the structure around the Dirac nodes. The dotted-white horizontal line shows the cross-section that we take in (g). (b) Berry curvature magnitude calculated with the tight-binding model in single-polarization limit with a frequency of $\omega_c=62.83$ THz to show that the qualitative features of the Berry curvature, and hence the Chern numbers do not change with the frequency. (c) Berry curvature magnitude calculated with the Dirac model in single-polarization limit with a frequency of $\omega_c=200$ THz for comparison with (a). (d) Berry curvature magnitude calculated with the tight-binding model for two-polarization model with frequency ratio $\gamma=1.2$ at a fundamental cavity frequency of $\omega_1=200$ THz. (e) Berry curvature magnitude calculated with the Dirac model for two-polarization model with frequency ratio $\gamma=1.2$ at a fundamental cavity frequency of $\omega_1=200$ THz for comparison with (d). (f) Focus around the Dirac nodes in the Brillouin zone in (d). (g) Cross-sections of the Berry curvature at $\mathbf{K}_x$, depicted as dotted-white line in (a), for $l=1-4$. Chern numbers of the bands are written at the top of the subfigures. The Berry phases due to the loops in the curvature are written on the figure, which leads to the numerically calculated Chern numbers. An effective mass of $m=0.02m_e$ is used for all figures.}
\label{Fig3}
\end{figure}

We first present the results in the single-polarization limit. Fig.~\ref{Fig3}(a) shows the magnitude of the Berry curvature, $\log|F_{l,xy}(k_x,k_y)|$, in the entire Brillouin zone calculated with the tight-binding model for band $l=4$. In order to resolve the loop structures around the Dirac node, we choose a sufficiently large cavity frequency $\omega_c=200$ THz, however the qualitative features of the Berry curvature does not change with the frequency. This can be seen in Fig.~\ref{Fig3}(b) which utilizes a cavity frequency of $\omega_c=62.83$ THz, and exhibits exactly the same number of bright loops around the bright points, which are the $\mathbf{K}$ and $\mathbf{K}'$ valleys. The bright white color denotes the dominant contribution to Berry curvature, whereas the dark black color is the minimum contribution. As a rule of thumb, we always observe $l-1$ bright closed loops around $\mathbf{K}/\mathbf{K'}$ valleys in Berry curvature for the $l^{\textrm{th}}$ band. Each of these closed loops carry Berry phase proportional to the number of chiral photon exchange with matter. This mechanism was introduced and shown in the main text. Here we provide extra evidence based on the tight-binding model results. Subfigures in Fig.~\ref{Fig3}(g) are the cross-sections at $\mathbf{K}_x$, white-dotted line (a), for four different photo-electron bands, $l=1-4$. The bright dots in the Berry curvature, (g$_1$), always carry either of $\pm \pi$ Berry phase originating from the pure matter degrees of freedom, Dirac nodes. Fig.~\ref{Fig3}(g$_2$) shows the cross-section for $l=2$ where two side-bands appear, each with $-\pi$ Berry phase contribution. Strictly speaking, these side-bands originate from the first closed loop around the Dirac node carrying a total of $-2\pi$ Berry phase due to a hybridization with a 1-photon exchange process. Fig.~\ref{Fig3}(g$_3$) shows the cross-section for $l=3$ where an additional two side-bands appear, however this time each with $-2\pi$ Berry phase contribution, because the light-matter 
hybridization occurs with a 2-photon exchange process. This corresponds to the second closed loop around the Dirac node. Finally looking at the band $l=4$, we observe the third loop around the Dirac node carrying a total of $-6\pi$ phase which translates to each side-band in (g$_4$) contributing $-3\pi$ Berry phase as depicted in the figure, due to a 3-photon exchange process. For a band $l=4$, this is the maximum number of closed loops that one would find in the Berry curvature due to the limit in the number of photon exchange processes. Finally let us point out Fig.~\ref{Fig3}(c) which plots the Berry curvature of a patch in the Brillouin zone calculated with the Dirac model. Remarkably, the Dirac model reproduces the exact physics of Berry curvature, giving rise to the correct counting of Berry phases. This is due to the fact that all light-matter hybridizations occur in the single-polarization limit exclusively around the high symmetry points of $\mathbf{K}$ and $\mathbf{K}'$. 

We also examine the Berry curvature for a two-polarization model with a frequency splitting ratio $\gamma=1.2$, which assumes a fundamental cavity frequency of $\omega_1=200$ THz in Figs.~\ref{Fig3}(d-f). The Berry curvature in this model follows very closely to the single-polarization limit, albeit the polarization of exchanged photons also plays a role, as explained in the main text. However, the Berry curvature still features closed loops around the Dirac nodes only. This is the reason why the Dirac model captures the essential physics as seen in Fig.~\ref{Fig3}(e). This figure should be contrasted to the focus on the Brillouin zone in Fig.~\ref{Fig3}(f), calculated with the tight-binding model. Let us note that the physics in this band, $l=4$, was already explained in great detail in the main text. This band experiences four total light-matter hybridizations including at the Dirac node. The innermost loop around the Dirac node is where two photons of different polarizations, so-to-speak, constructively interfere to give rise to $+4\pi$ Berry phase. The outer two loops occur due to simpler photon exchange processes where polarization does not directly play a role. Explicitly, the second and third loops occur due to a 1 left- and 2 right-circularly polarized photon exchanges, respectively. This can be easily checked with the band structure, Figs.~\ref{SFig1}(b)-(c).

\subsection{Dependence of vacuum gap on system parameters}

\begin{figure}
\centering
\includegraphics[width=0.85\columnwidth]{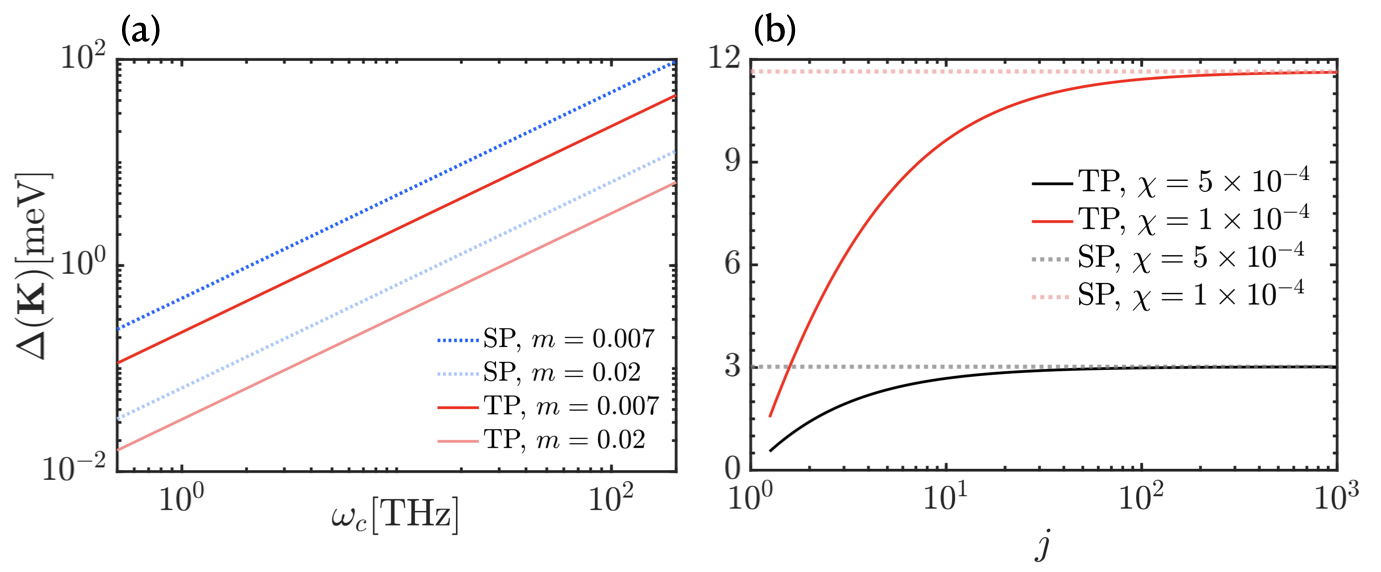}
\caption{Vacuum gap comparison between single-polarization and two-polarization models with a single Faraday rotator, calculated with the tight-binding formalism and exact diagonalization with a truncated photonic Hilbert space of maximum photon number $\langle a_{\lambda}^{\dagger}a_{\lambda}\rangle_{\textrm{max}}=4$. (a) Single-polarization (blue-dotted) and two-polarization (red-solid) for two different effective electron mass have been depicted with respect to cavity frequency. (b) Two-polarization vacuum gap (solid) is depicted with respect to the parameter $j$ in $\mathcal{V}_{L}= j \chi \left(2\pi c/ \omega_{c}\right)^3$ \cite{paravicini2019magneto} which is the effective cavity volume of left-circularly polarized light at an effective mass of $m=0.007$ \cite{Zhang_2005} for a cavity frequency of $\omega_c=6.28$ THz. Solid-black and -red are for different subwavelength cavity parameter $\chi$ which controls how strong the photons couple to electrons in the material. The dotted-lines are the single-polarization limit. As $j$ increases, and hence the left-circular polarized light couples to the material much weaker, two-polarization model approaches to the single-polarization limit in vacuum.}
\label{SFig2}
\end{figure}

We expand on the comparison between two-polarization model and its single-polarization limit in this section. Here we adopt the two-polarization model utilized in the discussion of vacuum gap. This model can be realized with a single Faraday rotator, and it only requires some amount of phase shift between left- and right-circular polarizations. We fix this ratio to be $g_R=\sqrt{2}g_L$, which has to be experimentally determined for a specific setup. Let us note that for clarity, one could also utilize the alternative two-polarization model where a frequency splitting is assumed, see previous subsections on the band structure. However, this does not change the physics, rather it simplifies the possible realization of this physics in the laboratory. 

Fig.~\ref{SFig2}(a) shows how increasing cavity frequency enhances the vacuum gap at the Dirac nodes for both single- and two-polarization models. Hence for instance utilizing a photonic crystal cavity \cite{doi:10.1021/acsphotonics.6b00219} could enhance the gap at least by an order of magnitude. Additionally Fig.~\ref{SFig2}(a) shows that lighter electrons exhibit a larger vacuum gap. Fig.~\ref{SFig2}(b) shows that engineering the effective cavity volume $\mathcal{V}= \chi \left(2\pi c/ \omega_{c}\right)^3$ \cite{paravicini2019magneto} through $ \chi$ parameter could boost the vacuum gap, as expected. This plot also shows for what value of $\chi_L=j\chi$ of left-circularly polarized light in an experimental setup, the single-polarization limit is still viable. One could see that $j > 10^2$, single-polarization model is a very good approximation. Importantly, the difference in the vacuum gap predicted by single- and two-polarization models are larger for a cavity with smaller cavity volume, and hence stronger light-matter interaction. In conclusion, the induced gap in graphene due to coupling to light could be used not only to determine whether the time-reversal symmetry is broken in the electromagnetic field, but also as a way to gauge the amount of phase shift between different polarizations of light.

\section{Derivation of effective Schrieffer–Wolff Hamiltonians}

We will apply Schrieffer–Wolff (SW) transformation~\cite{SWolf} to integrate out the photonic degrees of freedom in the lowest order appearing in the Hamiltonian. This transformation provides an effective Hamiltonian $H$ in the form of,
\begin{eqnarray}
    H = e^{S} \mathcal{H} e^{-S}, \notag
\end{eqnarray}
where $S$ is the transformation operator that satisfies the condition $[S,\mathcal{H}_0] = -\mathcal{H}_{\text{int}}$ in the decomposition $\mathcal{H}=\mathcal{H}_0+\mathcal{H}_{\text{int}}$ where $\mathcal{H}_{\text{int}}$ is the light-matter interaction Hamiltonian. With the condition satisfied, we obtain 
\begin{eqnarray}
     H = \mathcal{H}_0 + \frac{1}{2} [S,\mathcal{H}_{\text{int}}] + \mathcal{O}(g_R^3).  \notag
\end{eqnarray}
In the following we apply the SW transformation to the continuum model of graphene.

\subsection{Single-polarization model}

We assume $g_L=0$ for the single-polarization model as set in the main text. The following parts of the Hamiltonian include the Dirac model at $\mathbf{K}$ point, the cavity energy and the light-matter interaction,
\begin{eqnarray}
    \mathcal{H}_0 &=& \hbar v_F \sum_{\mathbf{k}} (k_x+ik_y)  c_{A\mathbf{k}}^{\dagger} c_{B\mathbf{k}} + \text{h.c.} + \hbar \Omega_R \left(a^{\dagger}_R a_R + \frac{1}{2}\right), \\
    \mathcal{H}_{\text{int}} &=& - \hbar v_F \sqrt{2} A_0 \sum_{\mathbf{k}} a^{\dagger}_R c_{A\mathbf{k}}^{\dagger} c_{B\mathbf{k}} + \text{h.c.}
\end{eqnarray}
Here we define $g_R\equiv \hbar \sqrt{2} A_0$, $\Omega_R \equiv \omega_R/v_F$ for convenience and assume $\hbar=1$, which recasts the equations to
\begin{eqnarray}
    \mathcal{H}_0 &=& \sum_{\mathbf{k}} (k_x+ik_y)  c_{A\mathbf{k}}^{\dagger} c_{B\mathbf{k}} + \text{h.c.} +  \Omega_R \left(a^{\dagger}_R a_R + \frac{1}{2}\right), \\
    \mathcal{H}_{\text{int}} &=& -g_R \sum_{\mathbf{k}} a^{\dagger}_R c_{A\mathbf{k}}^{\dagger} c_{B\mathbf{k}} + \text{h.c.}
\end{eqnarray}
We utilize the following commutators in the SW transformation,
\begin{eqnarray}
\left[c_{B\mathbf{k}}^{\dagger}c_{A\mathbf{k}},c_{A\mathbf{k}'}^{\dagger}c_{B\mathbf{k}'}\right] &=& \delta_{\mathbf{kk'}} \left(n_{B\mathbf{k}}-n_{A\mathbf{k}}\right),  \\
\left[c_{A\mathbf{k}}^{\dagger}c_{A\mathbf{k}},c_{A\mathbf{k}'}^{\dagger} c_{B\mathbf{k}'}\right] &=& \delta_{\mathbf{kk'}} c_{A\mathbf{k}}^{\dagger} c_{B\mathbf{k}}, \notag \\
\left[c_{A\mathbf{k}}^{\dagger}c_{A\mathbf{k}},c_{B\mathbf{k}'}^{\dagger} c_{A\mathbf{k}'}\right] &=&- \delta_{\mathbf{kk'}} c_{B\mathbf{k}}^{\dagger} c_{A\mathbf{k}},\notag \\
\left[ a_R  ,a^{\dagger}_R a_R \right] &=& a_R ,\notag \;\; \textrm{and}\;\; \left[  a^{\dagger}_R  ,a^{\dagger}_R a_R \right] = -  a^{\dagger}_R .\notag
\end{eqnarray}

The following transformation operator gives us the result up to the second order in perturbation theory, 
\begin{eqnarray}
    S= \tau_1\sum_{\mathbf{k}} \left( a^{\dagger}_R c_{A\mathbf{k}}^{\dagger} c_{B\mathbf{k}}  - a_R  c_{B\mathbf{k}}^{\dagger} c_{A\mathbf{k}}  \right) + \tau_2 \sum_{\mathbf{k}} \left[ -a^{\dagger}_R (k_x-ik_y) + a_R(k_x+ik_y) \right] (n_{A\mathbf{k}}-n_{B\mathbf{k}}),
\end{eqnarray}
with $\tau_1$ and $\tau_2$, such that $[S,\mathcal{H}_0]=-\mathcal{H}_{\text{int}}$. Both of these operators are expectantly anti-Hermitian. Let us note separately the commutator results,

\begin{eqnarray}
    [S_1,\mathcal{H}_0] &=& \tau_1 \sum_{\mathbf{k}} \bigg [ a^{\dagger}_R (k_x-ik_y)(n_{A\mathbf{k}}-n_{B\mathbf{k}}) - a_R  (k_x+ik_y)(-n_{A\mathbf{k}}+n_{B\mathbf{k}}) + \Omega_R \left( -a^{\dagger}_R  c_{A\mathbf{k}}^{\dagger} c_{B\mathbf{k}} - a_R  c_{B\mathbf{k}}^{\dagger} c_{A\mathbf{k}}  \right) \bigg],\nonumber\\
    \left[S_2,\mathcal{H}_0\right] &=& \tau_2\sum_{\mathbf{k}}\left(-a^{\dagger}_R(k_x-\textrm{i}k_y)+a_R (k_x+\textrm{i}k_y)\right)\left(2(k_x+\textrm{i}k_y)c^{\dagger}_{Ak}c_{Bk} -2(k_x-\textrm{i}k_y)c^{\dagger}_{Bk}c_{Ak}\right) \\
    &+& \tau_2 \Omega_R \sum_{\mathbf{k}} \left[ a^{\dagger}_R (k_x-ik_y) + a_R(k_x+ik_y) \right] (n_{A\mathbf{k}}-n_{B\mathbf{k}}).\notag
\end{eqnarray}
We find $\tau_1=-g_R/\Omega_R$ and $\tau_2=g_R/\Omega_R^2$ revealing that the small parameter in this perturbation theory is $1/\Omega_R$. Now we derive the effective Hamiltonian up to the second order. Let us start with,
\begin{eqnarray}
    [S,\mathcal{H}_{\textrm{int}}] &=& \bigg [ -\frac{g_R}{\Omega_R} \sum_{\mathbf{k}} \left( a^{\dagger}_R  c_{A\mathbf{k}}^{\dagger} c_{B\mathbf{k}}  - a_R  c_{B\mathbf{k}}^{\dagger} c_{A\mathbf{k}}  \right) +\frac{g_R}{\Omega_R^2} \sum_{\mathbf{k}} \left[ -a^{\dagger}_R (k_x-ik_y) + a_R(k_x+ik_y) \right] (n_{A\mathbf{k}}-n_{B\mathbf{k}}), \notag \\
    &-&g_R \sum_{\mathbf{k}} a^{\dagger}_R c_{A\mathbf{k}}^{\dagger} c_{B\mathbf{k}} + \text{h.c.}\bigg ].
\end{eqnarray}
To calculate this, we will make use of the following commutators
\begin{eqnarray}
    \left[a^{\dagger}_R c_{A\mathbf{k}}^{\dagger}c_{B\mathbf{k}}, a_R c_{B\mathbf{k}'}^{\dagger}c_{A\mathbf{k}'}\right] &=&  - c_{B\mathbf{k}'}^{\dagger}c_{A\mathbf{k}'}  c_{A\mathbf{k}}^{\dagger}c_{B\mathbf{k}} + a^{\dagger}_R a_R \hspace{1mm} \delta_{\mathbf{kk'}} \left(-n_{B\mathbf{k}}+n_{A\mathbf{k}}\right),\\
    \left[a_R c_{B\mathbf{k}}^{\dagger}c_{A\mathbf{k}},a^{\dagger}_R c_{A\mathbf{k}'}^{\dagger}c_{B\mathbf{k}'}\right] &=&  + c_{B\mathbf{k}}^{\dagger}c_{A\mathbf{k}}  c_{A\mathbf{k}'}^{\dagger}c_{B\mathbf{k}'} + a^{\dagger}_R a_R \hspace{1mm} \delta_{\mathbf{kk'}} \left(n_{B\mathbf{k}}-n_{A\mathbf{k}}\right).   \notag
\end{eqnarray}
This results in the first commutator, 
\begin{eqnarray}
[S_1,\mathcal{H}_{\textrm{int}}] &=& - \frac{g_R^2}{\Omega_R} \sum_{\mathbf{kk'} } c_{B\mathbf{k}}^{\dagger}c_{A\mathbf{k}}  c_{A\mathbf{k}'}^{\dagger}c_{B\mathbf{k}'} - 2\frac{g_R^2}{\Omega_R} a^{\dagger}_R a_R \sum_{\mathbf{k}}  \left(n_{B\mathbf{k}}-n_{A\mathbf{k}}\right) + \text{h.c.}
\end{eqnarray}
For the second commutator calculation, we need the following commutators,
\begin{eqnarray}
     \left[a^{\dagger}_R (n_{A\mathbf{k}}-n_{B\mathbf{k}}), a_R c_{B\mathbf{k}'}^{\dagger}c_{A\mathbf{k}'}\right] &=&- c_{B\mathbf{k}'}^{\dagger}c_{A\mathbf{k}'}(n_{A\mathbf{k}}-n_{B\mathbf{k}}) - 2 a^{\dagger}_R a_R \hspace{1mm} \delta_{\mathbf{kk'}} c_{B\mathbf{k}'}^{\dagger}c_{A\mathbf{k}'} \\
    \left[a_R (n_{A\mathbf{k}}-n_{B\mathbf{k}}), a^{\dagger} 
 c_{A\mathbf{k}'}^{\dagger}c_{B\mathbf{k}'}\right] &=& (n_{A\mathbf{k}}-n_{B\mathbf{k}}) c_{A\mathbf{k}'}^{\dagger} c_{B\mathbf{k}'} + 2 a^{\dagger}_R a_R \hspace{1mm} \delta_{\mathbf{kk'}} c_{A\mathbf{k}'}^{\dagger} c_{B\mathbf{k}'} \notag
\end{eqnarray}
Then the second commutator becomes,
\begin{eqnarray}
    \left[S_2,\mathcal{H}_{\textrm{int}} \right] &=& -\frac{g_R^2}{\Omega_R^2}  \sum_{\mathbf{kk'} } \bigg ( (k_x-ik_y) \left[ -a^{\dagger}_R (n_{A\mathbf{k}}-n_{B\mathbf{k}}), a^{\dagger}_R c_{A\mathbf{k}'}^{\dagger} c_{B\mathbf{k}'}   \right] + (k_x-ik_y) \left[ -a^{\dagger}_R(n_{A\mathbf{k}}-n_{B\mathbf{k}}), a_R \hspace{1mm} c_{B\mathbf{k}'}^{\dagger} c_{A\mathbf{k}'}   \right]  \notag \\
    &+& (k_x+ik_y) \left[ a_R (n_{A\mathbf{k}}-n_{B\mathbf{k}}), a^{\dagger}_R c_{A\mathbf{k}'}^{\dagger} c_{B\mathbf{k}'}  \right] + (k_x+ik_y) \left[ a_R \hspace{1mm} (n_{A\mathbf{k}}-n_{B\mathbf{k}}), a_R\hspace{1mm}c_{B\mathbf{k}'}^{\dagger} c_{A\mathbf{k}'}   \right]    \bigg ), \\
    &=& -\frac{g_R^2}{\Omega_R^2}  \sum_{\mathbf{kk'} } \bigg ( -2\left(a^{\dagger}_R\right)^2 (k_x-ik_y) \delta_{\mathbf{kk'}} c_{A\mathbf{k}'}^{\dagger} c_{B\mathbf{k}'} + (k_x-ik_y) \left( c_{B\mathbf{k}'}^{\dagger}c_{A\mathbf{k}'}(n_{A\mathbf{k}}-n_{B\mathbf{k}}) + 2 a^{\dagger}_R a_R \hspace{1mm} \delta_{\mathbf{kk'}} c_{B\mathbf{k}'}^{\dagger}c_{A\mathbf{k}'} \right) \notag \\
    &+& (k_x+ik_y) \left( (n_{A\mathbf{k}}-n_{B\mathbf{k}})c_{A\mathbf{k}'}^{\dagger}c_{B\mathbf{k}'} + 2 a^{\dagger}_R a_R \hspace{1mm} \delta_{\mathbf{kk'}}c_{A\mathbf{k}'}^{\dagger}c_{B\mathbf{k}'} \right) -2 a_R^2 (k_x+ik_y) \delta_{\mathbf{kk'}} c_{B\mathbf{k}'}^{\dagger} c_{A\mathbf{k}'}   \bigg ), \\
    &=& -\frac{g_R^2}{\Omega_R^2}  \sum_{\mathbf{kk'} } (k_x+ik_y) (n_{A\mathbf{k}}-n_{B\mathbf{k}})c_{A\mathbf{k}'}^{\dagger}c_{B\mathbf{k}'} + \frac{2g^2}{\Omega_R^2}  \sum_{\mathbf{k}} \left(  (k_x-ik_y) \left(a^{\dagger}_R\right)^2 - (k_x+ik_y) a^{\dagger}_R a_R  \right)c_{A\mathbf{k}}^{\dagger} c_{B\mathbf{k}} + \text{h.c.}\notag
\end{eqnarray}

Therefore, the entire effective Hamiltonian reads,
\begin{eqnarray}
H &=& \sum_{\mathbf{k}} \bigg[ k_x+ik_y +  \frac{g_R^2}{\Omega_R^2}  \left(  (k_x-ik_y) \left(a_R^{\dagger}\right)^2 - (k_x+ik_y) a^{\dagger}_R a_R  \right) + \mathcal{O}(g_R/\Omega_R^2,k^2,a_R,a_R^{\dagger}) \bigg]  c_{A\mathbf{k}}^{\dagger} c_{B\mathbf{k}} +  \Omega_R \left(a^{\dagger}_R a_R + \frac{1}{2}\right) \notag\\
&-&  \frac{g_R^2}{2\Omega_R}  \sum_{\mathbf{kk'} } \bigg (c_{B\mathbf{k}}^{\dagger}c_{A\mathbf{k}}  c_{A\mathbf{k}'}^{\dagger}c_{B\mathbf{k}'} +  \frac{k_x+ik_y}{\Omega_R} (n_{A\mathbf{k}}-n_{B\mathbf{k}})c_{A\mathbf{k}'}^{\dagger}c_{B\mathbf{k}'} \bigg) -  \frac{g_R^2}{\Omega_R} a^{\dagger}_R a_R \sum_{\mathbf{k}}  \left(n_{B\mathbf{k}}-n_{A\mathbf{k}}\right) + \text{h.c.}\label{eq:HeffSP}
\end{eqnarray}
Note that Eq.~\eqref{eq:HeffSP} reduces to the Eq.~(5) in the main text when it is projected to a cavity Fock state. The expression for the omitted term in the approximated SW transformation is,
\begin{eqnarray}
\mathcal{O}(g_R/\Omega_R^2,k^2,a_R,a^{\dagger}_R) &=& \frac{2g_R}{\Omega_R^2} \sum_{\mathbf{k}} \bigg[ - a^{\dagger}_R \left(k_x^2+k_y^2\right) + a_R \left(k_x^2 - k_y^2 + i 2k_x k_y \right) \bigg].
\end{eqnarray}
Since the Dirac model is valid only for infinitesimal $\mathbf{k}$ around $\mathbf{K}$ and $\mathbf{K'}$ points, it is valid to omit this term although its strength is $g_R/\Omega_R^2$. Nevertheless, as long as the cavity is in vacuum or in a Fock state with $\Braket{a_R^{\dagger}a_R}\in \mathbb{N}$, this term drops, and the effective SW Hamiltonian becomes exact. Let us conclude by making the remark that when the cavity is in vacuum state, $\Braket{a_R^{\dagger}a_R} = 0$, the effective Hamiltonian simplifies to,
\begin{eqnarray}
H_{\rm{vac}}&=& \sum_{\mathbf{k}} ( k_x+ik_y )  c_{A\mathbf{k}}^{\dagger} c_{B\mathbf{k}} +  \frac{\Omega_R}{2}-  \frac{g_R^2}{2\Omega_R}  \sum_{\mathbf{kk'} } \bigg (c_{B\mathbf{k}}^{\dagger}c_{A\mathbf{k}}  c_{A\mathbf{k}'}^{\dagger}c_{B\mathbf{k}'} +  \frac{k_x+ik_y}{\Omega_R} (n_{A\mathbf{k}}-n_{B\mathbf{k}})c_{A\mathbf{k}'}^{\dagger}c_{B\mathbf{k}'} \bigg)  + \text{h.c.}\label{eq:Heffvac}
\end{eqnarray}
Hamiltonian at $\mathbf{K}'$ valley can be similarly derived.

\subsection{Two-polarization model}

Focusing at $\mathbf{K}$ point with two polarization model, the non-interacting and light-matter interaction Hamiltonians follow as
\begin{eqnarray}
\mathcal{H}_0 &=& \sum_{\mathbf{k}} (k_x+ik_y)  c_{A\mathbf{k}}^{\dagger} c_{B\mathbf{k}} + \text{h.c.} +  \Omega_R \left(a_R^{\dagger}a_R + \frac{1}{2}\right) + \Omega_L \left(a_L^{\dagger}a_L + \frac{1}{2}\right), \\
\mathcal{H}_{\text{int}} &=& - \sum_{\mathbf{k}} \bigg(g_R a_R^{\dagger} + g_L a_L \bigg) c_{A\mathbf{k}}^{\dagger} c_{B\mathbf{k}} + \text{h.c.}
\end{eqnarray}
We choose the following SW transformation to find the effective Hamiltonian up to the first order, 
\begin{eqnarray}
    S= \sum_{\mathbf{k}} \left( \bigg[\alpha_1  a_R^{\dagger} - \alpha_2  a_L\bigg] c_{A\mathbf{k}}^{\dagger} c_{B\mathbf{k}}  - \bigg[\alpha_1  a_R - \alpha_2  a_L^{\dagger} \bigg] c_{B\mathbf{k}}^{\dagger} c_{A\mathbf{k}}  \right),
\end{eqnarray}
with $\alpha_1$ and $\alpha_2$, such that $[S,\mathcal{H}_0]= -\mathcal{H}_{\text{int}} + \mathcal{O}(g_{L(R)}/\Omega_{L(R)}^2)$. Note that we will truncate the SW for this calculation at the order of $\mathcal{O}(g_{L(R)}/\Omega_{L(R)}^2)$. The reason is the following: We already calculated the effective Hamiltonian for the single polarization model up to the second order in perturbation theory, and found out that the second order does not change the gap at the Dirac nodes, rather it perturbatively flattens the bands around the Dirac nodes, see Sec.~\ref{sec:MFTsecondOrder}. We find $\alpha_1=-g_R/\Omega_R$ and $\alpha_2=-g_L/\Omega_L$. We have
\begin{eqnarray}
[S,\mathcal{H}_{\textrm{int}}] &=& - \frac{g_R^2}{\Omega_R} \sum_{\mathbf{kk'}} c_{B\mathbf{k}}^{\dagger}c_{A\mathbf{k}}  c_{A\mathbf{k}'}^{\dagger}c_{B\mathbf{k}'} + \text{h.c.} - 2\frac{g_R^2}{\Omega_R} a_R^{\dagger} a_R \sum_{\mathbf{k}}  \left(n_{B\mathbf{k}}-n_{A\mathbf{k}}\right) \notag\\
&-& \frac{g_L^2}{\Omega_L} \sum_{\mathbf{kk'}} c_{A\mathbf{k}}^{\dagger}c_{B\mathbf{k}}  c_{B\mathbf{k}'}^{\dagger}c_{A\mathbf{k}'} +\text{h.c.}+ 2\frac{g_L^2}{\Omega_L} a_L^{\dagger} a_L \sum_{\mathbf{k}}  \left(n_{B\mathbf{k}}-n_{A\mathbf{k}}\right) \\
&+& g_R g_L\left(\frac{1}{\Omega_L}-\frac{1}{\Omega_R}\right) \left(a_R a_L + \textrm{h.c.}\right)  \left(n_{B\mathbf{k}}-n_{A\mathbf{k}}\right). \notag
\end{eqnarray}
Then the effective Hamiltonian reads,
\begin{eqnarray}\label{eq:fullEffHatK}
H_{\mathbf{K}} &=& \sum_{\mathbf{k}} \bigg[ k_x+ik_y + \mathcal{O}\bigg(\frac{g_{L(R)}^2}{\Omega_{L(R)}^2},k\bigg)+ \mathcal{O}\bigg(\frac{g_{L(R)}}{\Omega_{L(R)}^2},k^2\bigg) \bigg]  c_{A\mathbf{k}}^{\dagger} c_{B\mathbf{k}} +  \Omega_R \left(a_R^{\dagger}a_R + \frac{1}{2}\right)+\Omega_L \left(a_L^{\dagger}a_L + \frac{1}{2}\right) \label{eq:effHtwoPol}\\
&-&   \sum_{\mathbf{kk'}} \bigg ( \frac{g_R^2}{2\Omega_R} c_{B\mathbf{k}}^{\dagger}c_{A\mathbf{k}}  c_{A\mathbf{k}'}^{\dagger}c_{B\mathbf{k}'} +  \frac{g_L^2}{2\Omega_L} c_{A\mathbf{k}}^{\dagger}c_{B\mathbf{k}}  c_{B\mathbf{k}'}^{\dagger}c_{A\mathbf{k}'}  + \text{h.c.}\bigg) \notag \\
&+& \bigg(\frac{g_L^2}{\Omega_L} a_L^{\dagger} a_L -  \frac{g_R^2}{\Omega_R} a_R^{\dagger} a_R + \frac{g_R g_L}{2}\left(\frac{1}{\Omega_L}-\frac{1}{\Omega_R}\right) \left(a_R a_L + \textrm{h.c.}\right)  \bigg) \sum_{\mathbf{k}}  \left(n_{B\mathbf{k}}-n_{A\mathbf{k}}\right)\notag 
\end{eqnarray}
When the cavity is in a Fock state, $\Braket{a_R^{\dagger}a_R}\in \mathbb{N},\Braket{a_L^{\dagger}a_L} \in \mathbb{N}$, Eq.~\eqref{eq:effHtwoPol} simplifies to,
\begin{eqnarray} 
H_{\mathbf{K}} &=& v_F \sum_{\mathbf{k}} \bigg[ (k_x+ik_y)  c_{A\mathbf{k}}^{\dagger} c_{B\mathbf{k}} + \textrm{h.c.} + \bigg(\frac{g_L^2}{\Omega_L} a_L^{\dagger} a_L -  \frac{g_R^2}{\Omega_R} a_R^{\dagger} a_R  \bigg) \left(n_{B\mathbf{k}}-n_{A\mathbf{k}}\right) \bigg] \label{eq:fullEffTwoPolHatK}  \\
&-&  \sum_{\mathbf{kk'} } \bigg ( \frac{g_R^2 v_F}{2\Omega_R} c_{B\mathbf{k}}^{\dagger}c_{A\mathbf{k}}  c_{A\mathbf{k}'}^{\dagger}c_{B\mathbf{k}'} +  \frac{g_L^2v_F}{2\Omega_L} c_{A\mathbf{k}}^{\dagger}c_{B\mathbf{k}}  c_{B\mathbf{k}'}^{\dagger}c_{A\mathbf{k}'}  + \text{h.c.}\bigg) + \Omega_R \left(a_R^{\dagger}a_R + \frac{1}{2}\right)+\Omega_L \left(a_L^{\dagger}a_L + \frac{1}{2}\right)\notag
\end{eqnarray}
When the cavity is in vacuum state, $\Braket{a_R^{\dagger}a_R} = \Braket{a_L^{\dagger}a_L} = 0$, Eq.~\eqref{eq:effHtwoPol} simplifies to Eq.~(3) in the main text.

\section{Hartree-Fock mean-field theory}

\subsection{\label{sec:HFmeanfieldFirstOrder}Treating the first order interaction}

\begin{figure}
\centering
\includegraphics[width=1 \columnwidth]{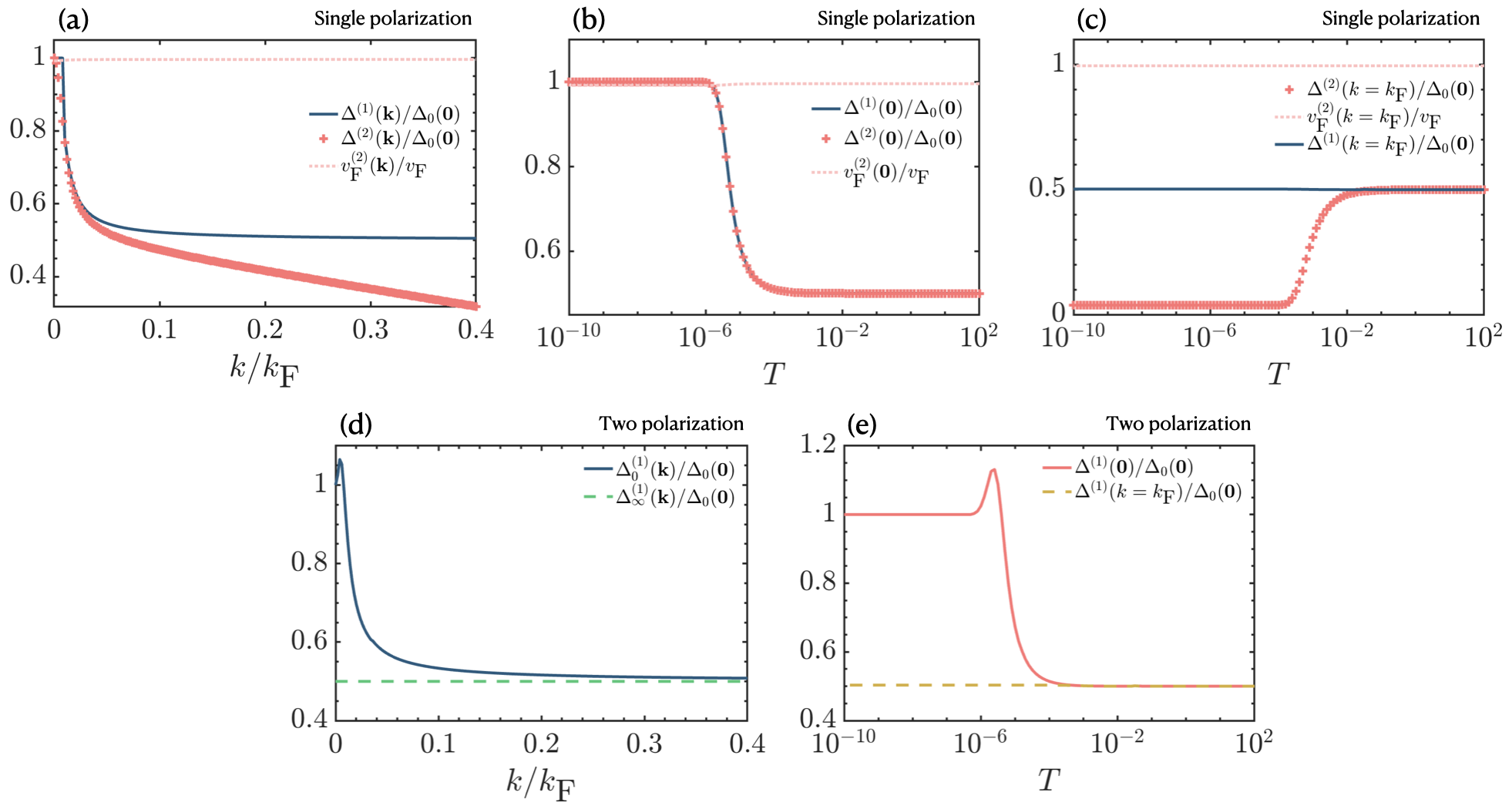}
\caption{Numerical solutions to the gap equations for single- (a-c) and two-polarization (d-e) models for fixed frequency $\omega_c=6.28$ THz and effective mass of $m=0.02m_e$. (a) Comparison of first and second order perturbation theories with respect to rescaled momentum $k=\sqrt{k_x^2+k_y^2}$ with Fermi momentum $k_{\textrm{F}}$ at zero temperature. (b) Comparison of first and second order perturbation theories with respect to temperature at the Dirac nodes. (c) Comparison of first and second order perturbation theories with respect to temperature at Fermi momentum $k_{\textrm{F}}$. In (a-c), the solid-blue and red-pluses are the induced gaps in the first and second order perturbation theories scaled with the zero temperature gaps at the Dirac nodes $\Delta_0(\mathbf{0})$, and the pink-dotted is the Fermi velocity renormalization. (d) First order perturbation theory with respect to rescaled momentum $k=\sqrt{k_x^2+k_y^2}$ with Fermi momentum $k_{\textrm{F}}$ at zero temperature (solid-blue) and infinite temperature (dashed-green). (e) First order perturbation theory with respect to temperature at Dirac node (solid-red) and Fermi momentum $k_{\textrm{F}}$ (dashed-yellow). In (d-e) we use a two-polarization model where only one Faraday rotator is needed which creates unequal couplings for left- and right-circularly polarized light $g_R=\sqrt{2}g_L$. }
\label{Fig4}
\end{figure}

Let us focus on the single-polarization model where $g_L=0$, and write Hartree and Fock terms by performing the Wick contractions, 
\begin{eqnarray}
    H_{1,\textrm{MFT}}^{\textrm{e-e}} &\sim & -  \frac{g_R^2}{\Omega_R}  \sum_{\mathbf{kk'}} \bigg(\Braket{ c_{B\mathbf{k}}^{\dagger}c_{A\mathbf{k}} }   c_{A\mathbf{k}'}^{\dagger}c_{B\mathbf{k}'}  +  c_{B\mathbf{k}}^{\dagger}c_{A\mathbf{k}} \Braket{ c_{A\mathbf{k}'}^{\dagger}c_{B\mathbf{k}'} }  - \Braket{ c_{B\mathbf{k}}^{\dagger}c_{A\mathbf{k}} }   \Braket{c_{A\mathbf{k}'}^{\dagger}c_{B\mathbf{k}'}}  \notag \\
    &+&  \Braket{ c_{B\mathbf{k}}^{\dagger} c_{B\mathbf{k}'}}  c_{A\mathbf{k}}  c_{A\mathbf{k}'}^{\dagger} +  c_{B\mathbf{k}}^{\dagger} c_{B\mathbf{k}'} \Braket{c_{A\mathbf{k}}  c_{A\mathbf{k}'}^{\dagger}}  - \Braket{ c_{B\mathbf{k}}^{\dagger} c_{B\mathbf{k}'}} \Braket{ c_{A\mathbf{k}}  c_{A\mathbf{k}'}^{\dagger}} \bigg). \label{eq:MFT1}
\end{eqnarray}
Note that strictly speaking, this is not an equality. Since we take two channels into account, we rescale Eq.~\eqref{eq:MFT1} by a factor of $2$,
\begin{eqnarray}
    H_{1,\textrm{MFT}}^{\textrm{e-e}} & \simeq & -  \frac{g_R^2}{2\Omega_R}  \sum_{\mathbf{kk'}} \bigg(\Braket{ c_{B\mathbf{k}}^{\dagger}c_{A\mathbf{k}} }   c_{A\mathbf{k}'}^{\dagger}c_{B\mathbf{k}'}  +  c_{B\mathbf{k}}^{\dagger}c_{A\mathbf{k}} \Braket{ c_{A\mathbf{k}'}^{\dagger}c_{B\mathbf{k}'} }  - \Braket{ c_{B\mathbf{k}}^{\dagger}c_{A\mathbf{k}} }   \Braket{c_{A\mathbf{k}'}^{\dagger}c_{B\mathbf{k}'}}  \notag \\
    &+&  \Braket{ c_{B\mathbf{k}}^{\dagger} c_{B\mathbf{k}'}}  c_{A\mathbf{k}}  c_{A\mathbf{k}'}^{\dagger} +  c_{B\mathbf{k}}^{\dagger} c_{B\mathbf{k}'} \Braket{c_{A\mathbf{k}}  c_{A\mathbf{k}'}^{\dagger}}  - \Braket{ c_{B\mathbf{k}}^{\dagger} c_{B\mathbf{k}'}} \Braket{ c_{A\mathbf{k}}  c_{A\mathbf{k}'}^{\dagger}} \bigg).\notag
\end{eqnarray}
Working with a thermal ensemble of electrons, we need to make sure that the Wick contractions are done with respect to an appropriate thermal ensemble which is the Fermi-Dirac distribution. This gives rise to the following \textit{nonzero} Wick contractions (see Sec.~\ref{sec:bog} for the intermediate steps),
\begin{eqnarray}
\Braket{c^{\dagger}_{A\mathbf{k}} c_{A\mathbf{k}}} &=& \frac{E_{\mathbf{k}} +\Delta(\mathbf{k})}{2E_{\mathbf{k}}} \frac{1}{e^{\beta E_{\mathbf{k}}}+1} + \frac{k_x^2+k_y^2}{2E_{\mathbf{k}}(E_{\mathbf{k}}+\Delta(\mathbf{k}))} \frac{1}{e^{-\beta E_{\mathbf{k}}}+1}= \frac{1}{2} \bigg( 1 - \frac{\Delta(\mathbf{k})\tanh (\beta E_{\mathbf{k}}/2)}{E_{\mathbf{k}}} \bigg), \label{eq:nA}\\
\Braket{c^{\dagger}_{B\mathbf{k}} c_{B\mathbf{k}}} &=& \frac{k_x^2+k_y^2}{2E_{\mathbf{k}}(E_{\mathbf{k}}+\Delta(\mathbf{k}))} \frac{1}{e^{\beta E_{\mathbf{k}}}+1} + \frac{E_{\mathbf{k}} +\Delta(\mathbf{k})}{2E_{\mathbf{k}}}\frac{1}{e^{-\beta E_{\mathbf{k}}}+1} = \frac{1}{2} \bigg( 1+ \frac{\Delta(\mathbf{k})\tanh (\beta E_{\mathbf{k}}/2)}{E_{\mathbf{k}}} \bigg),\label{eq:nB} \\
\Braket{c^{\dagger}_{A\mathbf{k}} c_{B\mathbf{k}}} &=&  \frac{k_x-ik_y}{2E_{\mathbf{k}}} \bigg(\frac{1}{e^{\beta E_{\mathbf{k}}}+1} -  \frac{1}{e^{-\beta E_{\mathbf{k}}}+1} \bigg) = -\frac{k_x-ik_y}{2E_{\mathbf{k}}}  \tanh\bigg(\frac{\beta E_{\mathbf{k}}}{2} \bigg).\label{eq:AB}
\end{eqnarray}
For now, let us only consider the channel regarding the Eqs.~\eqref{eq:nA} and~\eqref{eq:nB}, because these channels are responsible for the gap opening. The remaining channel's effect is shown to be zero in Sec.~\ref{sec:secondChannel}. Then, we obtain an MFT Hamiltonian in the first order
\begin{eqnarray}
H^{(1)}_{\textrm{MFT}}  &=&\sum_{\mathbf{k}}\bigg[ ( k_x+ik_y)  c_{A\mathbf{k}}^{\dagger} c_{B\mathbf{k}} + \text{h.c.} - \frac{g_R^2}{2\Omega_R} \bigg( \Braket{ c_{B\mathbf{k}}^{\dagger} c_{B\mathbf{k}}} (- c_{A\mathbf{k}}^{\dagger}  c_{A\mathbf{k}} +  c_{B\mathbf{k}}^{\dagger} c_{B\mathbf{k}})  - \Braket{ c_{B\mathbf{k}}^{\dagger} c_{B\mathbf{k}}}^2 +\Braket{ c_{B\mathbf{k}}^{\dagger} c_{B\mathbf{k}}}\bigg) \bigg].  \notag \\
&=& \sum_{\mathbf{k}}\bigg[ v_F( k_x \sigma_1 + k_y \sigma_2 ) -  \frac{g_R^2 v_F^2}{2\omega_R} \Braket{ c_{B\mathbf{k}}^{\dagger} c_{B\mathbf{k}}} \sigma_3 \bigg] + E_0. \label{eq:int1MFT2}
\end{eqnarray}
Here we make a definition for the gap,
\begin{eqnarray}
\Delta(\mathbf{k}) &=&  \frac{g_R^2 v_F^2}{2\omega_R} \Braket{ c_{B\mathbf{k}}^{\dagger} c_{B\mathbf{k}}},   
\end{eqnarray}
as this is the amplitude of $\sigma_3$ term. $E_0$ in Eq.~\eqref{eq:int1MFT2} is the many-body ground state energy,
\begin{eqnarray}
    E_0 &=& \frac{\omega_R}{2}+\frac{g_R^2v_F^2}{2\omega_R} \sum_{\mathbf{k}} \bigg(\Braket{ c_{B\mathbf{k}}^{\dagger} c_{B\mathbf{k}}}^2 -\Braket{ c_{B\mathbf{k}}^{\dagger} c_{B\mathbf{k}}}\bigg) = \frac{\omega_R}{2}+ \sum_{\mathbf{k}} \bigg(\frac{2\omega_R}{g_R^2 v_F^2}\Delta(\mathbf{k})^2 -\Delta(\mathbf{k})\bigg).
\end{eqnarray}
By using the Wick rotations, we find the gap equation to be
\begin{eqnarray}
\Delta(\mathbf{k}) &=& \frac{g_R^2 v_F^2}{2\omega_R} \frac{1}{2} \bigg( 1+ \frac{\Delta(\mathbf{k})\tanh (\beta E_{\mathbf{k}}/2)}{E_{\mathbf{k}}} \bigg).\label{eq:gapEq}
\end{eqnarray}
Let us note that at zero temperature and exactly at $\mathbf{K}$ point, we obtain $\Delta_0(\mathbf{0})=g_R^2v_F^2/2\omega_R$ as already discussed in the Letter. Here the notation $\Delta_T(\mathbf{k})$ denotes the vacuum gap at temperature $T$ and momentum $\mathbf{k}$. In the opposite limit where the inverse temperature $\beta \rightarrow 0$, we can expand $\tanh$ around $0$, and obtain
\begin{eqnarray}
 \Delta_{\infty}(\mathbf{0}) &=& \frac{g_R^2 v_F^2}{4\omega_R} +  \frac{g_R^2 v_F^2}{4\omega_R}  \frac{\Delta}{E_{\mathbf{k}}} \bigg(\frac{\beta}{2}E_{\mathbf{k}} - \frac{1}{3}\left(\frac{\beta}{2}E_{\mathbf{k}} \right)^3 +\mathcal{O}(\beta^5) \bigg) \\
 &=& \frac{g_R^2 v_F^2}{4\omega_R} +  \frac{g_R^2 v_F^2}{2^3\omega_R}  \Delta\beta - \frac{g_R^2 v_F^2}{3\times2^5\omega_R}  \Delta E_{\mathbf{k}}^2 \beta^3 +\mathcal{O}(\beta^5)= \frac{\frac{g_R^2 v_F^2}{4\omega_R}}{1-\frac{g_R^2 v_F^2}{8\omega_R}\beta}. \notag
\end{eqnarray}
If $\beta \ll \frac{8\omega_R}{g_R^2 v_F^2}$, we obtain
\begin{eqnarray}
 \Delta_{\infty}(\mathbf{0}) &=&  \frac{g_R^2 v_F^2}{4\omega_R}.  
\end{eqnarray}
Hence we find, the gap based on the first order perturbation, changes from $\frac{g_R^2 v_F^2}{2\omega_R}$ to $\frac{g_R^2 v_F^2}{4\omega_R}$ as temperature increases. Except these temperature points, we cannot solve the equation analytically. Hence we solve Eq.~\eqref{eq:gapEq} numerically to find the dependence on the temperature. Fig.~\ref{Fig4}(a-c) depicts the numerical results to the gap equation in the first order perturbation theory in the single-polarization limit (blue-solid). In (a) we observe that as we move away from the Dirac nodes, the gap decreases. Fig.~\ref{Fig4}(b) shows how the gap decreases with increasing temperature. Fig.~\ref{Fig4}(c) shows that the gap at the Fermi momentum is not affected by the temperature. 

\subsection{\label{sec:bog}Bogoliubov diagonalization for the MFT Hamiltonian in the single-polarization limit}

Here we give the exact diagonalization of the Hamiltonian
\begin{eqnarray}
    H &=& \sum_{\mathbf{k}} \bigg( k_x \sigma_1 + k_y \sigma_2 - \Delta(\mathbf{k}) \sigma_3 + E_0(\mathbf{k}) \bigg).
\end{eqnarray}
This form appears after the Wick contractions. Since this is a fermionic system, the Bogoliubov transformation follows as
\begin{eqnarray}
\gamma_{\alpha\mathbf{k}} = u c_{A\mathbf{k}} + v c_{B\mathbf{k}}, \;\; \textrm{and}\;\;
\gamma_{\xi\mathbf{k}} = - v^* c_{A\mathbf{k}} + u^* c_{B\mathbf{k}},
\end{eqnarray}
while the inverse transformation is
\begin{eqnarray}
c_{A\mathbf{k}} = u^* \gamma_{\alpha\mathbf{k}}  - v \gamma_{\xi\mathbf{k}},\; \; \textrm{and}\;\;
c_{B\mathbf{k}} = v^* \gamma_{\alpha\mathbf{k}}+ u\gamma_{\xi\mathbf{k}}.
\end{eqnarray}
To write $H$ in terms of the new basis, we need to find
\begin{eqnarray}
c^{\dagger}_{A\mathbf{k}} c_{A\mathbf{k}} &=& |u|^2 \gamma^{\dagger}_{\alpha\mathbf{k}} \gamma_{\alpha\mathbf{k}} + |v|^2 \gamma^{\dagger}_{\xi\mathbf{k}} \gamma_{\xi\mathbf{k}} - uv \gamma^{\dagger}_{\alpha\mathbf{k}} \gamma_{\xi\mathbf{k}} -u^*v^* \gamma^{\dagger}_{\xi\mathbf{k}} \gamma_{\alpha\mathbf{k}}, \\
c^{\dagger}_{B\mathbf{k}} c_{B\mathbf{k}} &=& |v|^2 \gamma^{\dagger}_{\alpha\mathbf{k}} \gamma_{\alpha\mathbf{k}} + |u|^2 \gamma^{\dagger}_{\xi\mathbf{k}} \gamma_{\xi\mathbf{k}} + uv \gamma^{\dagger}_{\alpha\mathbf{k}} \gamma_{\xi\mathbf{k}} + u^*v^* \gamma^{\dagger}_{\xi\mathbf{k}} \gamma_{\alpha\mathbf{k}}, \notag \\
c^{\dagger}_{A\mathbf{k}} c_{B\mathbf{k}} &=& uv^* \gamma^{\dagger}_{\alpha\mathbf{k}} \gamma_{\alpha\mathbf{k}} -  uv^* \gamma^{\dagger}_{\xi\mathbf{k}} \gamma_{\xi\mathbf{k}} +u^2 \gamma^{\dagger}_{\alpha\mathbf{k}} \gamma_{\xi\mathbf{k}} -(v^*)^2 \gamma^{\dagger}_{\xi\mathbf{k}} \gamma_{\alpha\mathbf{k}}. \notag
\end{eqnarray}
Substituting these into the Hamiltonian results in the following three terms,
\begin{eqnarray}
    H_0 &=& \sum_{\mathbf{k}} E_0(\mathbf{k}). \\
    H_1 &=& \sum_{\mathbf{k}} \bigg[(k_x + ik_y )uv^* + \text{h.c.} - \Delta(\mathbf{k}) \bigg( |v|^2 - |u|^2 \bigg) \bigg] \bigg(\gamma_{\alpha\mathbf{k}}^{\dagger}\gamma_{\alpha\mathbf{k}} - \gamma_{\xi\mathbf{k}}^{\dagger}\gamma_{\xi\mathbf{k}}\bigg)\\
    H_2 &=& \sum_{\mathbf{k}} \bigg[(k_x + ik_y )u^2 - (k_x - ik_y ) v^2 - 2uv \Delta(\mathbf{k}) \bigg] \gamma_{\alpha\mathbf{k}}^{\dagger}\gamma_{\xi\mathbf{k}}+ \text{h.c.}  
\end{eqnarray}
For diagonalization to happen, $H_2=0$ which leads to the expressions for $u$ and $v$. By also using the fact that $|u|^2+|v|^2=1$,
\begin{eqnarray}
    (k_x + ik_y )u^2 - (k_x - ik_y ) v^2 - 2uv \Delta(\mathbf{k}) &=& 0 \rightarrow u = \sqrt{\frac{E_{\mathbf{k}} +\Delta(\mathbf{k})}{2E_{\mathbf{k}}}}, \hspace{2mm} v=\frac{k_x+ik_y}{\sqrt{2E_{\mathbf{k}}(E_{\mathbf{k}}+\Delta(\mathbf{k}))}},
\end{eqnarray}
where $E_{\mathbf{k}}^2=k_x^2+k_y^2+\Delta(\mathbf{k})^2$ is the excitation energy. Substituting these into $H_1$, we obtain
\begin{eqnarray}
    H_1 &=& \sum_{\mathbf{k}} E_{\mathbf{k}} \bigg(\gamma_{\alpha\mathbf{k}}^{\dagger}\gamma_{\alpha\mathbf{k}} - \gamma_{\xi\mathbf{k}}^{\dagger}\gamma_{\xi\mathbf{k}}\bigg).
\end{eqnarray}
Hence we see that the lower energy band is denoted by $\beta$ with energy $-E_{\mathbf{k}}$. Note that the Bogoliubov quasi-particles (Bogoliubons) follow the Fermi-Dirac distribution when Wick contraction is applied in Sec.~\ref{sec:HFmeanfieldFirstOrder}
\begin{eqnarray}
    \Braket{\gamma_{\alpha\mathbf{k}}^{\dagger}\gamma_{\alpha\mathbf{k}} } = \frac{1}{e^{\beta E_{\mathbf{k}}}+1}, \hspace{5mm}
 \Braket{\gamma_{\xi\mathbf{k}}^{\dagger}\gamma_{\xi\mathbf{k}} }=\frac{1}{e^{-\beta E_{\mathbf{k}}}+1},   
\end{eqnarray}
where $\beta$ is the inverse temperature. The tunneling terms are
\begin{eqnarray}
\Braket{\gamma_{\alpha\mathbf{k}}^{\dagger}\gamma_{\xi\mathbf{k}} } &=& \text{Tr} \bigg[\rho \gamma_{\alpha\mathbf{k}}^{\dagger}\gamma_{\xi\mathbf{k}}  \bigg] =\text{Tr} \bigg[ \sum_{\sigma\mathbf{k}} \frac{\gamma_{\sigma\mathbf{k}}^{\dagger} \Ket{\mathbf{0}}\Bra{\mathbf{0}} \gamma_{\sigma\mathbf{k}}}{e^{\beta E_{\sigma}(\mathbf{k})}-1}  \gamma_{\alpha\mathbf{k}}^{\dagger} \gamma_{\xi\mathbf{k}}  \bigg] =0.
\end{eqnarray}

\subsection{\label{sec:secondChannel}Considering the second channel}

Here let us discuss the fate of the channel $\sum_{\mathbf{k}} \Braket{c^{\dagger}_{A\mathbf{k}} c_{B\mathbf{k}}}$. We immediately observe the following equation
\begin{eqnarray}
v_F \kappa_c &=& -\frac{g_R^2 v_F^2}{\omega_R} \sum_{\mathbf{k'}} \Braket{c^{\dagger}_{A\mathbf{k}'} c_{B\mathbf{k}'}} = \frac{g_R^2 v_F^2}{\omega_R} \sum_{\mathbf{k'}} \frac{v_F (k'_x-ik'_y + \kappa_c)}{2E_{\mathbf{k'}}}  \tanh\bigg(\frac{\beta E_{\mathbf{k'}}}{2}\bigg) \in \mathbb{C}. \notag
\end{eqnarray}
First, at infinite temperature $\beta\rightarrow 0$, $\kappa_c=0$. For the other temperatures, this gives an integral equation over the Fermi surface
\begin{eqnarray}
\kappa_c &=& \frac{g_R^2 v_F^2}{\omega_R} \int_{-\kappa_y}^{\kappa_y} dk'_y   \int_{-\kappa_x}^{\kappa_x} dk'_x \frac{k'_x-ik'_y + \kappa_c}{2E_{\mathbf{k'}}}  \tanh\bigg(\frac{\beta E_{\mathbf{k'}}}{2}\bigg) \notag
\end{eqnarray}
Note that $\kappa_c$ is momentum independent, by definition. Also since $\kappa_c$ can be written as a mere shift on the momenta,
\begin{eqnarray}
\kappa_c &=& \frac{g_R^2 v_F^2}{\omega_R} \int_{-\kappa_y}^{\kappa_y} dk'_y   \int_{-\kappa_x}^{\kappa_x} dk'_x \frac{(k'_x+ \kappa_c^x) - i (k'_y + \kappa_c^y)}{2E_{\mathbf{k'}}}  \tanh\bigg(\frac{\beta E_{\mathbf{k'}}}{2}\bigg)   \notag
\end{eqnarray}
We do a change of variables, $k_x = k_x'+\kappa_c^x$ and $k_y = k_y'+\kappa_c^y$,
\begin{eqnarray}
\kappa_c &=& \frac{g_R^2 v_F^2}{\omega_R} \int_{-\kappa_y+\kappa_c^y}^{\kappa_y+\kappa_c^y} dk_y   \int_{-\kappa_x+\kappa_c^x}^{\kappa_x+\kappa_c^x} dk_x \frac{k_x - i k_y}{2E_{\mathbf{k}}}  \tanh\bigg(\frac{\beta E_{\mathbf{k}}}{2}\bigg).   \notag
\end{eqnarray}
Now let us change to polar coordinates, $k=\sqrt{k_x^2+k_y^2}$ and $\tan \phi = k_y/k_x$,
\begin{eqnarray}
\kappa_c &=& \frac{g_R^2 v_F^2}{\omega_R} \int_{-\pi}^{\pi} d\phi \int_{|\kappa_c|}^{|\kappa_c|+\kappa_{\textrm{F}}}k dk  \frac{e^{-i\phi}}{2E_{k}}  \tanh\bigg(\frac{\beta E_{k}}{2}\bigg) = 0. \notag
\end{eqnarray}
where $|\kappa_c| = |\kappa_c^x-i\kappa_c^y|=\sqrt{(\kappa_c^x)^2+(\kappa_c^y)^2}$, and $\kappa_{\textrm{F}} = \sqrt{\kappa_x^2+\kappa_y^2}$ is the Fermi momentum. Hence
we find $\kappa_c=0$, and $\sum_{\mathbf{k}}\Braket{c^{\dagger}_{A\mathbf{k}} c_{B\mathbf{k}}}=0$ at any temperature. Hence the first line of Eq.~\eqref{eq:MFT1} vanishes.

\subsection{\label{sec:MFTsecondOrder}MFT on the second order interaction term}

In this section, we perform the MFT to the second order interaction term in Eq.~\eqref{eq:Heffvac}. By using the following expressions, 
\begin{eqnarray}
 \frac{g_R^2}{2\Omega_R^2}  (k_x+ik_y) \Braket{  c_{B\mathbf{k}}^{\dagger}c_{B\mathbf{k}} } &=&   \frac{g_R^2}{4\Omega_R^2}  e^{i\phi}  k  \bigg( 1 + \frac{\Delta(\mathbf{k})\tanh (\beta E_{\mathbf{k}}/2)}{E_{\mathbf{k}}} \bigg), \label{eq:MFTsecD1} \\
\frac{g_R^2}{2\Omega_R^2}  \sum_{\mathbf{k}} (k_x+ik_y) \Braket{  c_{A\mathbf{k}}^{\dagger}c_{A\mathbf{k}} } &=&   \frac{g_R^2}{2\Omega_R^2}   \frac{1}{4\pi \kappa_{\textrm{F}}} \int_{-\pi}^{\pi} e^{i\phi} d\phi  \int_{0}^{\kappa_{\textrm{F}}} k^2 dk \bigg( 1 - \frac{\Delta(\mathbf{k})\tanh (\beta E_{\mathbf{k}}/2)}{E_{\mathbf{k}}} \bigg) = 0, \notag \\
-  \frac{g_R^2}{2\Omega_R^2}  \sum_{\mathbf{k}} (k_x+ik_y) \Braket{  c_{B\mathbf{k}}^{\dagger}c_{B\mathbf{k}} } &=& 0.\notag
\end{eqnarray}
we find, 
\begin{eqnarray}
H_{2,\textrm{MFT}}^{\textrm{e-e}} &=&  -  \frac{g_R^2}{4\Omega_R^2}  \sum_{\mathbf{k} } (k_x+ik_y) \bigg ( -  \Braket{ c_{A\mathbf{k}}^{\dagger} c_{B\mathbf{k}}} c_{A\mathbf{k}}^{\dagger} c_{A\mathbf{k}}  +  c_{A\mathbf{k}}^{\dagger} c_{B\mathbf{k}} \Braket{c_{B\mathbf{k}}^{\dagger} c_{B\mathbf{k}} } - \Braket{ c_{A\mathbf{k}}^{\dagger} c_{B\mathbf{k}}} \Braket{c_{B\mathbf{k}}^{\dagger} c_{B\mathbf{k}} } \notag \\
&+& \Braket{ c_{B\mathbf{k}}^{\dagger} c_{B\mathbf{k}}} c_{A\mathbf{k}}^{\dagger} c_{B\mathbf{k}} + c_{B\mathbf{k}}^{\dagger} c_{B\mathbf{k}} \Braket{ c_{A\mathbf{k}}^{\dagger} c_{B\mathbf{k}}}-\Braket{c_{B\mathbf{k}}^{\dagger} c_{B\mathbf{k}}} \Braket{c_{A\mathbf{k}}^{\dagger} c_{B\mathbf{k}} } + \Braket{ c_{A\mathbf{k}}^{\dagger} c_{B\mathbf{k}}} \bigg) + \text{h.c.}    
\end{eqnarray}
See the Sec.~\ref{sec:secondChannel} for the definition of $\phi$ and $k$ appeared in Eqs.~\eqref{eq:MFTsecD1}. Hartree-Fock expansion leads to the following MFT Hamiltonian in the second order perturbation theory,
\begin{eqnarray}
 H_{\textrm{MFT}}^{(2)} &=& \sum_{\mathbf{k} }\bigg[ \bigg( 1-\frac{g_R^2}{2\Omega_R^2}\Braket{c_{B\mathbf{k}}^{\dagger} c_{B\mathbf{k}}}  \bigg)( k_x+ik_y)  c_{A\mathbf{k}}^{\dagger} c_{B\mathbf{k}} + \text{h.c.} \label{eq:MFT2ndOrder} \\
 &-& \bigg( \frac{g_R^2}{2\Omega_R}\Braket{c_{B\mathbf{k}}^{\dagger} c_{B\mathbf{k}}}+(k_x+ik_y) \frac{g_R^2}{4\Omega_R^2} \Braket{c_{A\mathbf{k}}^{\dagger}c_{B\mathbf{k}} } + \text{h.c.} \bigg) (- c_{A\mathbf{k}}^{\dagger}  c_{A\mathbf{k}} +  c_{B\mathbf{k}}^{\dagger} c_{B\mathbf{k}}) \notag \\
 &+& (k_x+ik_y) \frac{g_R^2}{4\Omega_R^2} \Braket{c_{A\mathbf{k}}^{\dagger}c_{B\mathbf{k}} } \bigg(2\Braket{c_{B\mathbf{k}}^{\dagger} c_{B\mathbf{k}}} - 1\bigg) +\text{h.c.} +\frac{g_R^2}{2\Omega_R} \Braket{c_{B\mathbf{k}}^{\dagger} c_{B\mathbf{k}}}^2 -\frac{g_R^2}{2\Omega_R}\Braket{c_{B\mathbf{k}}^{\dagger} c_{B\mathbf{k}}} \bigg]. \notag
\end{eqnarray}
Hence we have a new gap equation (with Fermi velocity $v_F$ substituted back) and an equation for the Fermi velocity renormalization,
\begin{eqnarray}
\Delta^{(2)}(\mathbf{k})&=& \frac{g_R^2v_F^2}{2\omega_R}\Braket{c_{B\mathbf{k}}^{\dagger} c_{B\mathbf{k}}}+ v_F (k_x+ik_y) \frac{g_R^2v_F^2}{4\omega_R^2} \Braket{c_{A\mathbf{k}}^{\dagger}c_{B\mathbf{k}} } + \text{h.c.}   \label{eq:MFT2ndgap1} \\
v_F^{(2)}(\mathbf{k}) &=& \bigg( 1-\frac{g_R^2v_F^2}{2\omega_R^2}\Braket{c_{B\mathbf{k}}^{\dagger} c_{B\mathbf{k}}}  \bigg)v_F. \label{eq:MFTFermiVel1}
\end{eqnarray}
Let us substitute Eqs.~\eqref{eq:MFT2ndgap1} and~\eqref{eq:MFTFermiVel1} into Eq.~\eqref{eq:MFT2ndOrder},
\begin{eqnarray}
 H_{\textrm{MFT}}^{(2)} &=& \sum_{\mathbf{k} }\bigg[v_F^{(2)}(\mathbf{k}) ( k_x+ik_y)  c_{A\mathbf{k}}^{\dagger} c_{B\mathbf{k}} + \text{h.c.}- \Delta^{(2)}(\mathbf{k})  (- c_{A\mathbf{k}}^{\dagger}  c_{A\mathbf{k}} +  c_{B\mathbf{k}}^{\dagger} c_{B\mathbf{k}})\bigg] + E_0^{(2)},
\end{eqnarray}
where $E_0^{(2)}$  is the many-body ground state energy in the second order perturbation theory,
\begin{eqnarray}
    E_0^{(2)} &=& \frac{\omega_R}{2}+ \sum_{\mathbf{k}} \bigg[ - \Delta^{(2)}(\mathbf{k}) + \frac{4\omega_R^2}{g_R^2v_F^2}\Delta^{(2)}(\mathbf{k}) \left( 1- \frac{v_F^{(2)}(\mathbf{k})}{v_F}\right) - \frac{2\omega_R^3 }{g_R^2v_F^2} \bigg(1-\frac{v_F^{(2)}(\mathbf{k})}{v_F}\bigg)^2 \bigg].
\end{eqnarray}
The energy expression in the second order reads $E_{\mathbf{k}}^2 = (v_F^{(2)}(\mathbf{k})k)^2+(\Delta^{(2)}(\mathbf{k}))^2$ noting $v_F^{(2)}(\mathbf{k})\in \mathbb{R}$. The Wick contractions should be updated as $v_F k_x \rightarrow v_F^{(2)}(\mathbf{k})k_x$, $v_F k_y \rightarrow v_F^{(2)}(\mathbf{k})k_y$ and $\Delta(\mathbf{k})\rightarrow \Delta^{(2)}(\mathbf{k})$ below.
\begin{eqnarray}
\Braket{c^{\dagger}_{A\mathbf{k}} c_{A\mathbf{k}}} &=& \frac{1}{2} \bigg( 1 - \frac{\Delta^{(2)}(\mathbf{k})\tanh (\beta E_{\mathbf{k}}/2)}{E_{\mathbf{k}}} \bigg), \label{eq:nA2} \\
\Braket{c^{\dagger}_{B\mathbf{k}} c_{B\mathbf{k}}} &=& \frac{1}{2} \bigg( 1+ \frac{\Delta^{(2)}(\mathbf{k})\tanh (\beta E_{\mathbf{k}}/2)}{E_{\mathbf{k}}} \bigg),\notag \\
\Braket{c^{\dagger}_{A\mathbf{k}} c_{B\mathbf{k}}} &=&  -v_F^{(2)}(\mathbf{k})\frac{k_x-ik_y}{2E_{\mathbf{k}}}  \tanh\bigg(\frac{\beta E_{\mathbf{k}}}{2} \bigg).\notag
\end{eqnarray}
The coupled MFT equations for the second order perturbation theory follow after Eqs.~\eqref{eq:nA2} are substituted into Eqs.~\eqref{eq:MFT2ndgap1} and~\eqref{eq:MFTFermiVel1},
\begin{eqnarray}
\Delta^{(2)}(\mathbf{k})&=& \frac{g_R^2v_F^2}{4\omega_R} \bigg( 1+ \frac{\Delta^{(2)}(\mathbf{k})\tanh (\beta E_{\mathbf{k}}/2)}{E_{\mathbf{k}}} \bigg) - v_F^{(2)}(\mathbf{k}) v_F k^2 \frac{g_R^2v_F^2}{2\omega_R^2} \frac{\tanh(\beta E_{\mathbf{k}}/2)}{2E_{\mathbf{k}}}  \\
v_F^{(2)}(\mathbf{k}) &=& v_F -\frac{g_R^2 v_F^3}{4\omega_R^2 }\bigg( 1+ \frac{\Delta^{(2)}(\mathbf{k})\tanh (\beta E_{\mathbf{k}}/2)}{E_{\mathbf{k}}} \bigg) .\notag
\end{eqnarray}
Let us state these coupled MFT equations at $\mathbf{K}$ point,
\begin{eqnarray}
\Delta^{(2)}( \mathbf{0})&=& \frac{g_R^2v_F^2}{4\omega_R} \bigg( 1+ \tanh (\beta \Delta^{(2)}(\mathbf{0})/2) \bigg)  \\
v_F^{(2)}(\mathbf{0}) &=& v_F \bigg(1-\frac{\Delta^{(2)}(\mathbf{0})}{2\omega_R} \bigg) .\notag
\end{eqnarray}
The gap equation turns out to be the same with the gap equation in the first order perturbation theory, Eq.~\eqref{eq:gapEq}. However, the second order perturbation introduces a renormalization to the Fermi velocity. At zero temperature, $\beta\rightarrow \infty$, equations can be analytically solved
\begin{eqnarray}
\Delta^{(2)}_0(\mathbf{0})= \frac{g_R^2 v_F^2}{2\omega_R} \;\;\; \textrm{and}\;\;\;
v_{F,0}^{(2)}(\mathbf{0}) = v_F \bigg(1-\frac{g_R^2v_F^2}{2\omega_R^2} \bigg). \label{eq:vFermiNormAtzero}
\end{eqnarray}
We solve the equations numerically around $\mathbf{K}$ point in terms of Fermi momentum $\kappa_{\textrm{F}}$.  

Fig.~\ref{Fig4}(a-c) summarizes the numerical solutions to the gap equation in both first and second order perturbation theories in the single-polarization limit. In (a), we observe that Fermi velocity renormalization is negligible at the Dirac nodes and away from them. This is consistent within the perturbation theory, see Eq.~\eqref{eq:vFermiNormAtzero}. At the Dirac nodes, the effect is the largest and is practically independent of the temperature, Fig.~\ref{Fig4}(b). However the gap in the second order perturbation theory drastically differs from the first order as we move away from the Dirac nodes. This has an important effect in the match with the numerical band structure as shown in the main text, Fig.~1. The gap in both orders is the same at the Dirac nodes independently from the temperature, Fig.~\ref{Fig4}(b), and it expectantly decreases with the temperature. We observe that the gap at the Fermi momentum increases with temperature in the second order, eventually converges to the same value with the first order gap at large temperatures, Fig.~\ref{Fig4}(c).

\subsection{MFT on two-polarization model}

There are two interaction terms in the effective SW Hamiltonian of two-polarization model, Eq.~\eqref{eq:fullEffTwoPolHatK}, and they have to be treated separately giving rise to two independent gap equations. These interactions give rise to the following effective MFT Hamiltonians,
\begin{eqnarray}
    H^{(R)}_{\textrm{MFT}} =  -\sum_{\mathbf{k}} \frac{g_R^2 v_F^2}{2\omega_R} \Braket{ c_{B\mathbf{k}}^{\dagger} c_{B\mathbf{k}}} \sigma_3 \;\;\; \textrm{and}\;\;\;
H^{(L)}_{\textrm{MFT}} = \sum_{\mathbf{k}} \frac{g_L^2 v_F^2}{2\omega_L}  \Braket{ c_{A\mathbf{k}}^{\dagger} c_{A\mathbf{k}}}  \sigma_3, 
\end{eqnarray}
where $H_0 = v_F( k_x \sigma_1 + k_y \sigma_2 )$ in both cases. Then defining 
\begin{eqnarray}
\Delta_R(\mathbf{k}) =  \frac{g_R^2 v_F^2}{2\omega_R} \Braket{ c_{B\mathbf{k}}^{\dagger} c_{B\mathbf{k}}} \;\;\; \textrm{and} \;\;\;
\Delta_L(\mathbf{k}) = \frac{g_L^2 v_F^2}{2\omega_L}  \Braket{ c_{A\mathbf{k}}^{\dagger} c_{A\mathbf{k}}},\notag
\end{eqnarray}
we can write
\begin{eqnarray}
    H_{\mathbf{K}}^{\textrm{MFT}} &=& \sum_k \bigg( k_x \sigma_1 + k_y \sigma_2 - \bigg[\Delta_R(\mathbf{k}) - \Delta_L(\mathbf{k}) \bigg] \sigma_3 + E_0(\mathbf{k}) \bigg).
\end{eqnarray}
By using Eqs.~\eqref{eq:nA} and~\eqref{eq:nB}, we find the gap equations stated in the Letter, Eq.~(4). Total gap is, $\Delta(\mathbf{k}) = \Delta_R(\mathbf{k}) - \Delta_L(\mathbf{k})$. Then one could see, $\Delta(\mathbf{k})=0$ always holds if the time reversal symmetry is preserved. The many-body ground state energy will follow closely to the ones found for single polarization model in this case. Let us work this out explicitly. We have $E_0(\mathbf{k})= E_{0,R}(\mathbf{k}) + E_{0,L}(\mathbf{k})$ with,
\begin{eqnarray}
    E_{0,R} = \frac{\omega_R}{2}+ \sum_k \bigg(\frac{2\omega_R}{g_R^2 v_F^2}\Delta_R(\mathbf{k})^2 -\Delta_R(\mathbf{k})\bigg) \;\;\; \textrm{and}\;\;\;
E_{0,L} = \frac{\omega_L}{2}+ \sum_k \bigg(\frac{2\omega_L}{g_L^2 v_F^2}\Delta_L(\mathbf{k})^2 -\Delta_L(\mathbf{k})\bigg).   \notag 
\end{eqnarray}
For higher energy gaps at the Dirac nodes, the result follows very similar to the single polarization case too:
\begin{eqnarray}
    \Delta_0(\mathbf{0}) &=& (1+2n_R)\frac{g_R^2 v_F^2}{2\omega_R} - (1+2n_L)\frac{g_L^2 v_F^2}{2\omega_L}.
\end{eqnarray}

Figs.~\ref{Fig4}(d-e) show the numerically extracted gap for the two-polarization model where $g_R=\sqrt{2}g_L$ is set due to the phase shift anticipated with the Faraday effect. Both with respect to momentum and temperature, we observe very similar behaviors to the single-polarization limit. Let us note that the calculated gaps are rescaled with $\Delta_{0}(\mathbf{0})$, the gap at zero temperature and Dirac nodes in the two-polarization model considered. 

\end{document}